\documentclass[preprint,review,12pt]{elsarticle}

\usepackage{lineno}
\usepackage{graphicx}
\usepackage{url}
\usepackage{comment}
\usepackage{xcolor}
\usepackage[output-decimal-marker={.}]{siunitx}
\usepackage[version=4]{mhchem}
\DeclareSIUnit\barn{b}
\DeclareSIUnit\ton{t}
\DeclareSIUnit\angstrom{\text{Å}}
\DeclareSIUnit\bar{bar}
\usepackage{adjustbox}
\usepackage{float}
\usepackage{caption}
\usepackage{subcaption}
\usepackage{amsmath} 
\usepackage{amsfonts}
\usepackage{fancyhdr}
\usepackage{csquotes}
\usepackage{graphicx}
\usepackage{tikz}
\usepackage{listings}
\usepackage{wrapfig}
\usepackage{xcolor}
\usepackage{comment}
\usepackage{siunitx}
\usepackage{booktabs}
\usepackage{enumitem}
\usepackage{ulem}
\usepackage{rotating}
\usepackage{tabularx}

\begin{document}

\begin{frontmatter}

\title{ 
Application of LHC Gas Recuperation Systems for Methane Emission Control in Livestock Housing}

\author[a,b]{Ilaria Vai\corref{cor1}} 
\author[a,b]{Francesco Alessandro Angiulli}
\author[b]{Chiara Aimè}
\author[c,d]{Maria Cristina Arena}
\author[e]{Davide Biagini}
\author[d]{Amin Bouzaiene}
\author[b]{Alessandro Braghieri}
\author[a,b]{Matteo Brunoldi}
\author[b]{Domenico Calabrò}
\author[g]{Simone Calzaferri}
\author[a,b]{Alessandro Caserio}
\author[e]{Elio Dinuccio}
\author[c,b]{Daniele Dondi}
\author[f]{Linda Finco}
\author[a,b]{Giulia Giannandrea}
\author[b]{Samuel Guelfo Gigli}
\author[a,b]{Gabriele Giunta}
\author[d]{Roberto Guida}
\author[a]{Nithish Kumar Kameswaran}
\author[d]{Beatrice Mandelli}
\author[a,b]{Paolo Montagna}
\author[a,b]{Cristina Riccardi}
\author[b]{Paola Salvini}
\author[b]{Claudio Scagliotti}
\author[a,b]{Alessandro Tamigio}
\author[c]{Dhanalakshmi Vadivel}
\author[b]{Filippo Vercellati}
\author[f]{Riccardo Verna}
\author[a,b]{Paolo Vitulo}

\affiliation[a]{organization={Dipartimento di Fisica ``Alessandro Volta'', Università di Pavia},
            addressline={Via Bassi~6}, 
            city={Pavia},
            postcode={27100}, 
            country={Italy}}
\affiliation[b]{organization={Istituto Nazionale di Fisica Nucleare, Sezione di Pavia},
            addressline={Via Bassi~6}, 
            city={Pavia},
            postcode={27100}, 
            country={Italy}}
\affiliation[c]{organization={Dipartimento di Chimica, Università di Pavia},
            addressline={Via Torquato Taramelli~12}, 
            city={Pavia},
            postcode={27100}, 
            country={Italy}}
\affiliation[d]{organization={CERN},
            addressline={Esplanade des Particules~1}, 
            city={Geneva},
            postcode={1211}, 
            country={Switzerland}}
\affiliation[e]{organization={Dipartimento di Scienze Agrarie, Forestali e Alimentari, Università di Torino},
            addressline={Largo Paolo Braccini~2}, 
            city={Grugliasco (TO)},
            postcode={10095}, 
            country={Italy}}
\affiliation[f]{organization={Istituto Nazionale di Fisica Nucleare, Sezione di Torino},
            addressline={Via Pietro Giuria~1}, 
            city={Torino},
            postcode={10125}, 
            country={Italy}}

\affiliation[g]{organization={Department of Physics, Yonsei University},
            addressline={Science Hall 327A, 50 Yonsei-ro, Seodaemun-gu}, 
            city={Seoul},
            postcode={03722}, 
            country={South Korea}}

\cortext[cor1]{Corresponding author. \textit{Email address:} ilaria.vai@unipv.it} 

\begin{abstract}

The CH4rLiE (CH4 Livestock Emission) project investigates the technical feasibility of adapting gas recovery systems from high-energy physics to mitigate methane (\ce{CH4}) emissions in livestock housing. This work presents a proof-of-principle based on the adaptation of CERN’s gas recuperation systems for the capture of \ce{CH4} at low concentrations. A laboratory-scale prototype was developed to evaluate the performance of various adsorbent materials under realistic conditions, including multi-stage humidity removal and pressurized gas flows. Experimental results obtained with the prototype led to the selection of commercial Z5 zeolite as the primary adsorbent due to its high adsorption capacity and stable regeneration performance through Vacuum Swing Adsorption cycles. The study demonstrates the feasibility of \ce{CH4} capture at concentrations down to $0.1\%$. Furthermore, it was observed that increasing the \ce{CH4} partial pressure enhances the adsorption capacity, with tests conducted up to approximately \SI{5}{bar}. To bridge the gap between laboratory conditions and the representative $10-\SI{100}{ppm}$ levels found in dairy barn environments, a negative exponential extrapolation was applied to the experimental data. This allowed for the modeling of the adsorption behavior in the ultra-low concentration regime. These results validate the operational principle and provide the necessary parameters for the design of a full-scale system for field installation.

\end{abstract}

\end{frontmatter}

\section{Introduction}
\label{sec1}

The CH4 Livestock Emission (CH4rLiE) project aims to adapt gas recovery systems, originally designed for particle detectors, to methane (\ce{CH4}) capture, bridging the gap between high-energy physics technology and environmental mitigation.

As the most abundant greenhouse gas (GHG) in the troposphere after water vapor and carbon dioxide (\ce{CO2}), \ce{CH4} concentration significantly influences the Earth’s radiative balance~\cite{wuebbels}. Since the 1700s, human activities have more than doubled \ce{CH4} emissions, leading to atmospheric concentrations that continue to increase steadily. Because \ce{CH4} has a high global warming potential (GWP) yet is a short-lived gas,, reducing its atmospheric load offers an effective short-term strategy to mitigate global warming and improve air quality.

A major challenge lies in anthropogenic activities where \ce{CH4} is emitted in low concentrations over extended periods. A primary example is animal husbandry: in a typical dairy barn, concentrations may remain below \SI{100}{ppm}, yet a single cow can emit approximately \SI{110}{kg} of \ce{CH4} annually --- equivalent to \SI{3080}{kg} of \ce{CO2}. Globally, enteric fermentation and manure management account for 15\% of annual \ce{CH4} emissions~\cite{fao}.

The CH4rLiE project operates within this specific technical window, seeking to validate whether recovery systems developed for gaseous detector mixtures at the Large Hadron Collider (LHC)~\cite{guida} can be effectively repurposed for \ce{CH4} capture. While these technologies were developed to manage several pollutant components, the starting point for this adaptation is the system optimized for carbon tetrafluoride (\ce{CF4}), an F-gas (fluorinated greenhouse gas) used in Cathode Strip Chambers (CSC) with a GWP over a 100-year horizon ($\text{GWP}_{100}$) of 7390~\cite{AR6_IPCC}. Due to its high cost and environmental impact --- accounting for roughly 20\% of CERN’s direct GHG emissions~\cite{Dati_ricircolo_CF4} --- CERN has developed a four-pillar strategy to manage \ce{CF4} based on:

\begin{enumerate}
    \item gas recirculation systems;
    \item gas recuperation systems;
    \item research into novel, environmentally friendly gases for particle detectors;
    \item development of treatment plants to decompose GHGs into harmless byproducts.
\end{enumerate}

This study focuses exclusively on the first two areas. The recovery and recirculation of gases in these systems rely fundamentally on the use of adsorbent materials to selectively separate the target gas from the mixture~\cite{Y}. While the original CERN systems were optimized for \ce{CF4}, the same principle can be applied to \ce{CH4} by selecting appropriate materials. Numerous types of adsorbents are widely used for \ce{CH4} capture from both high- and low-concentration sources, including zeolites, activated carbon, metal-organic frameworks, carbon molecular sieves (MS), and ionic liquids~\cite{Y1}. The efficiency of these materials --- particularly critical when isolating \ce{CH4} from low-concentration streams --- is determined by their physical characteristics and pore structure, which encompasses morphology, pore size distribution, connectivity, volume, and specific surface area~\cite{Y1, X}. 

Based on these properties, various commercial materials were evaluated to determine their suitability for methane adsorption, leading to the selection of commercial zeolites for this study. Accordingly, the following sections discuss the operational principles of this technology, the development of adsorbent materials optimized for \ce{CH4}, the results of laboratory tests performed on a capture system prototype, and the final design of a system intended for installation in livestock barns. A comprehensive discussion of the remaining CERN strategies is available in Ref.~\cite{Dati_ricircolo_CF4}.

\section{LHC gas recirculation and recuperation systems}

The active medium in gas detectors consists of gas mixtures enclosed within a region where an electric field is generally present. As ionizing particles traverse the gas, they generate primary ionization, and the resulting signals are collected by dedicated electrodes. To mitigate the accumulation of space charge and impurities in high-rate applications --- such as experiments at the LHC --- the gas is typically circulated through the detectors. The effluent mixture, referred to as ``exhausted'', primarily comprises the original components along with potential impurities. Consequently, through appropriate filtration, these components can be reclaimed and integrated back into the fresh mixture supplied to the detectors.

\begin{figure}[!h]
    \centering
    \includegraphics[width=0.8\linewidth]{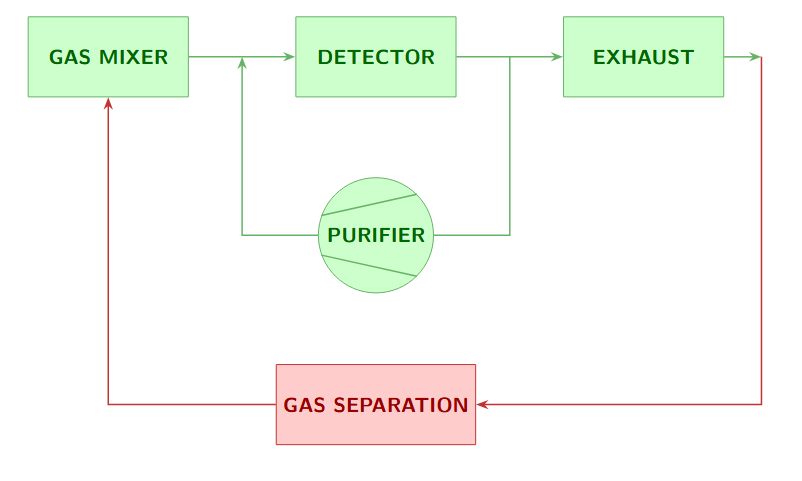}
    \caption{Schematic view of a recirculation system for a gaseous detector (in green), with the addition of recuperation system (in red). Adapted from Ref.~\cite{Tesi_MCA}}. 
    \label{fig:schema_ricircolo}
\end{figure}

Gas recirculation systems are therefore designed to facilitate the reuse of exhausted gases following its circulation through the detectors. As illustrated in green in Fig.~\ref{fig:schema_ricircolo}, the gas is collected upon exiting the active volume, purified, and re-injected into the detector system. 
In principle, a recirculation fraction of up to 99\% is achievable, depending upon the degree of mixture contamination. A primary limiting factor is the accumulation of impurities such as nitrogen (\ce{N2}), which are not fully eliminated by standard filtration processes. At the LHC, most gas systems currently achieve recirculation rates exceeding 90\%~\cite{CF4_recuperation}. However, the mandatory injection of the remaining 10\% fresh gas mixture constitutes a persistent and non-negligible source of GHG emissions.

In contrast, gas recuperation systems are designed to recover high-value or environmentally detrimental gases from the exhaust stream. In red in Fig.~\ref{fig:schema_ricircolo}, the effluent gas from the detectors is routed to a dedicated separation plant that isolates and extracts specific target components. Compared to standard recirculation, recuperation systems enable a near-total reduction in gas loss by not only purifying the components but also selectively harvesting them from the exhausted stream. This approach has played a pivotal role in mitigating GHG emissions since 2012, following the commissioning of the \ce{CF4} recuperation system.
Within this facility, the primary gas mixture, which accumulates  contaminants such as \ce{O2} and \ce{N2}, is channeled through an extraction line to separate its constituents. The separation process employs highly selective porous materials, namely zeolites, which adsorb one or more components of the mixture while permitting the others to flow downstream~\cite{Application_zeolites_environmental}.

\subsection{The \ce{CF4} recuperation system}

\begin{figure}[!h]
    \centering
    \includegraphics[width=\linewidth]{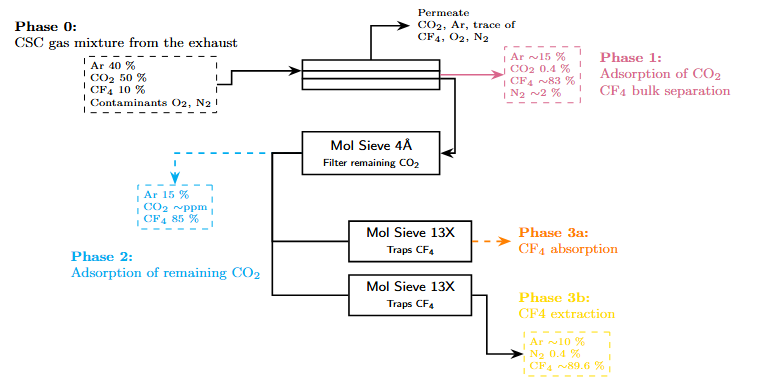}
    \caption{Process flow of the \ce{CF4} recuperation system for CSCs. Adapted from Ref.~\cite{Tesi_MCA}.}
    \label{fig:schema_recupero}
\end{figure}

This system is employed for the recovery of the CSCs gas mixture and comprises several components designed to efficiently isolate the target gas, as shown in Fig.~\ref{fig:schema_recupero}. The initial gas mixture consists of approximately 50\% \ce{CO2}, 40\% \ce{Ar}, and 10\% \ce{CF4}. Upon exiting the detectors, the effluent gas --- now carrying contaminants such as \ce{O2} and \ce{N2} --- is routed into the recovery line. The objective is the sequential separation of all mixture components, via the following specialized modules:
\begin{enumerate}
    \item a membrane module for bulk \ce{CO2} separation;
    \item a \SI{4}{\angstrom} MS for the removal of residual \ce{CO2};
    \item a 13X MS for the selective absorption of \ce{CF4};
    \item a \ce{CF4} compressor and corresponding storage tanks.
\end{enumerate}
The membrane module is engineered to extract the bulk of the \ce{CO2} from the mixture. However, due to imperfect selectivity, minor fractions of \ce{CF4} may be lost with the \ce{CO2} stream, while some \ce{CO2} remains in the primary gas flow. To eliminate this residual \ce{CO2}, the mixture is channeled through a subsequent module containing \SI{4}{\angstrom} zeolites (pore diameter $\sim$\SI{0.4}{\nano m}). These zeolites selectively adsorb \ce{CO2}, permitting \ce{CF4} to bypass in the next phase. This stage is crucial because the downstream 13X zeolite module (pore size $\sim$\SI{1}{\nano m}) is capable of adsorbing both \ce{CO2} and \ce{CF4}. The main differences between zeolite types are discussed in Sec.~\ref{confronto_zeoliti}. Consequently, the upstream removal of \ce{CO2} is necessary to maximize \ce{CF4} selectivity and recovery yield. \ce{Ar} and \ce{N2} exhibit negligible interaction with the adsorbents, occupying the free volume of the module ($\sim$60-70\%). Following adsorption, the \ce{CF4} is desorbed via material regeneration, typically achieved through temperature increase or pressure reduction. The resulting mixture of \ce{CF4}, \ce{Ar}, and \ce{N2} is subsequently compressed into a dedicated storage volume for re-usage into the detector gas mixture.

\begin{figure}[!h]
\centering
\adjustbox{center}{
\begin{tabular}{c c}

\includegraphics[width=0.3\textwidth]{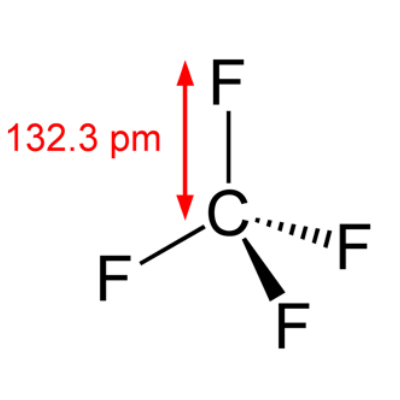} &
\includegraphics[width=0.3\textwidth]{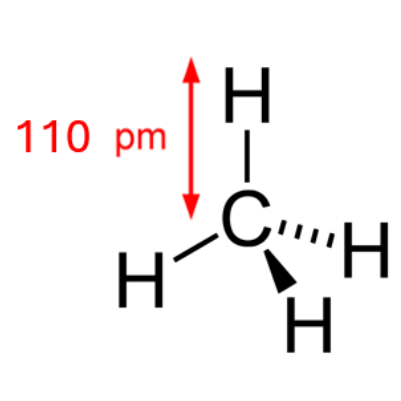} \\
(a) & (b) \\
\end{tabular}
}
\caption{Molecular structure of \ce{CF4} (a) and \ce{CH4} (b). }
\label{fig3:CF4_and_CH4_structure}
\end{figure}

The methane recovery system developed within the framework of the CH4rLiE project operates on analogous principles. Given that both \ce{CH4} and \ce{CF4} are tetrahedral molecules with similar molecular dimensions (Fig.~\ref{fig3:CF4_and_CH4_structure}), it is reasonable to assume that adsorbents effective for \ce{CF4} capture will similarly demonstrate high adsorption capacities for \ce{CH4}.

\section{Preliminary considerations on dedicated filter materials}
As introduced in the previous section, the starting point of the analysis of adsorbent materials dedicated to methane capture is the observation that both \ce{CH4} and \ce{CF4} are tetrahedral molecules with similar molecular dimensions (Fig.~\ref{fig3:CF4_and_CH4_structure}). For this reason, it is reasonable to begin the analysis of filter materials with commercial zeolites, particularly Z5, which are widely used in capture systems and which, as discussed in Section~\ref{sec:lab_proto}, will also be employed in the prototype developed in this study.

Zeolite Socony Mobil-5 (ZSM-5 or simply Z5) was first patented by Mobil Oil in 1975 and has since been extensively studied for gas adsorption. 
\begin{figure}[!h]
    \centering
    \includegraphics[width=0.5\linewidth]{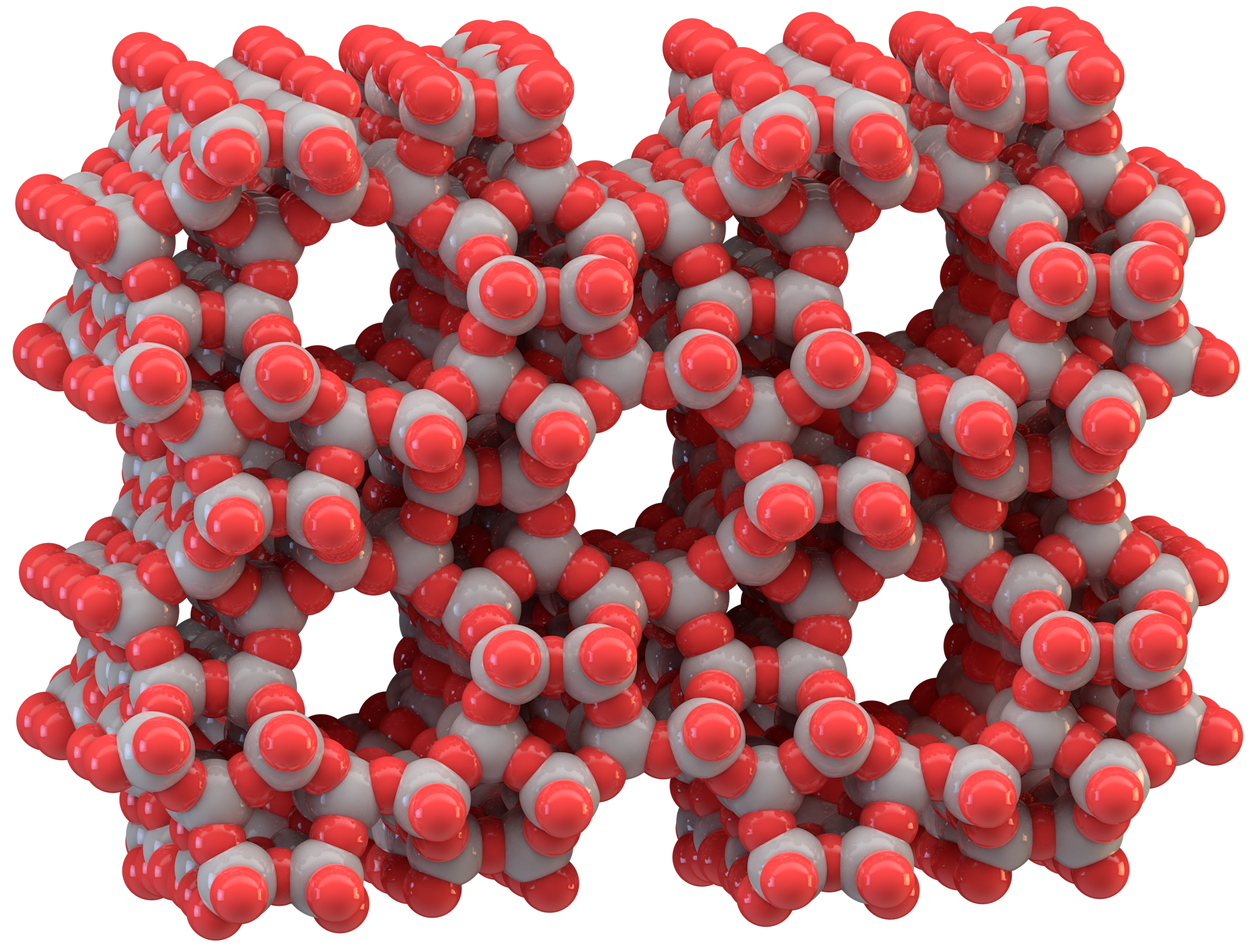}
    \caption{The microporous molecular structure of a zeolite, ZSM-5 or Z5. This aluminosilicate zeolite has a high silicon and low alumininum content~\cite{z5_wikicommons}.}
    \label{fig:Z5_structure}
\end{figure}
Z5 (Fig.~\ref{fig:Z5_structure}) features a Mobil Five (MFI) framework topology, which is built by connecting pentasil silicate chains. This structure forms a system of intersecting 10-membered ring channel: one sinusoidal channel with a pore size of $5.1\times \SI{5.5}{\angstrom}$ along the [100] axis, and one straight channel with a pore size of $5.3\times \SI{5.6}{\angstrom}$ along the [010] as described in Ref.~\cite{X1}. The effective pore opening size of Z5 (\SI{5}{\angstrom}) exceeds the kinetic diameter of methane (\SI{3.8}{\angstrom}), allowing it to readily accommodate methane molecules during the adsorption process~\cite{X2}. Furthermore, Z5 is well-regarded for its high chemical and thermal stability~\cite{X3}. The phase purity and crystallinity of the commercial Z5 molecular sieves analyzed in the context of this project were examined using Powder X-Ray Diffraction (PXRD) and compared with reference patterns from the International Zeolite Association (IZA) (Fig.~\ref{fig1_chim})~\cite{X4}. 
\begin{figure}
    \centering
    \includegraphics[width=0.7\textwidth]{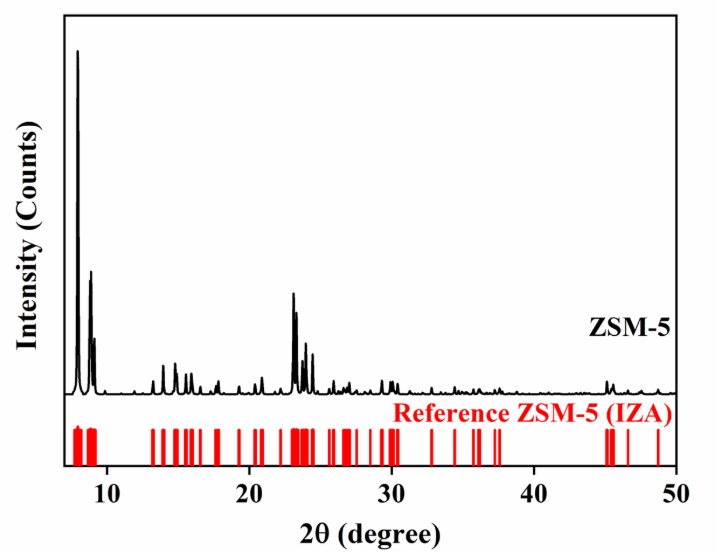}
    \caption{Comparison of PXRD patterns of commercial Z5 (black) and reference patterns obtained from the IZA (red). }
    \label{fig1_chim}
\end{figure}
\begin{figure}
    \centering
    \includegraphics[width=0.8\textwidth]{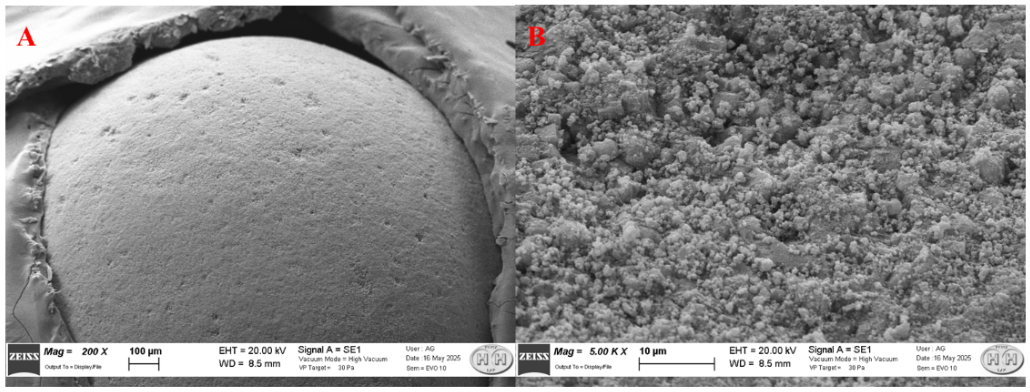}
    \caption{(A) SEM image of Z5 molecular sieve at 200x magnification. (B) SEM image of Z5 molecular sieve at 5000x magnification.}
    \label{fig2_chim}
\end{figure}
The absence of broad peaks confirms that the Z5 sample is highly crystalline, and the data are in excellent agreement with the benchmark given by IZA~\cite{X4}. Scanning Electron Microscope (SEM) images at 200x and 5000x magnifications (Fig.~\ref{fig2_chim} A and B, respectively) reveal a densely aggregated assembly of microcrystalline particles. These particles exhibit the characteristic irregular, granular morphology associated with MFI-type zeolites. The crystallites are tightly packed, with numerous visible interparticle voids. This irregular granular morphology provides a high surface area conducive for methane adsorption. Furthermore, SEM-Energy Dispersive X-ray Spectroscopy (SEM-EDX) was performed to determine the elemental composition and verify chemical purity; the results are summarized in Table~\ref{table_chim}. 
\begin{table}[ht]
    \centering
    \caption{Elemental analysis of Z5 zeolite with SEM-EDX.}
    \label{tab:composizione_elementi}
    \begin{tabular}{l S[table-format=2.2] S[table-format=2.2]}
        \toprule
        Element & {Weight Percentage (\%)} & {Atomic Percentage (\%)} \\
        \midrule
        O  & 54.47 & 68.06 \\
        Na & 3.77  & 3.28  \\
        Al & 14.11 & 10.46 \\
        Si & 19.02 & 13.54 \\
        Ca & 6.13  & 3.06  \\
        \bottomrule
    \end{tabular}
    \label{table_chim}
\end{table}
The Si/Al ratio, calculated with the atomic percentages, is 1.29, indicating an aluminum-rich environment. Additionally, \ce{Na+} and Ca$^{2+}$ cations were detected, which likely act as charge-balancing ions. The presence of these cations is highly beneficial, as they induce electrostatic interactions that polarize guest molecules, irrespective of the molecule's inherent polarity~\cite{X5}. This polarization effect is particularly  advantageous for trapping non-polar methane molecules. In summary, the characterizations confirm that the Z5 proposed is highly crystalline, possesses a high surface area with a favorable 3D channel system, and contains cations capable of polarizing non-polar methane. Combined with its well-documented scalability, tunability, and robust thermochemical stability~\cite{X3}, Z5 fulfills the requirements of an effective adsorbent and was therefore considered for this study~\cite{X6}.       

\begin{figure}[htbp]
\centering
\begin{subfigure}{0.50\textwidth}
  \centering
  \includegraphics[height=5cm, width=\linewidth, keepaspectratio]{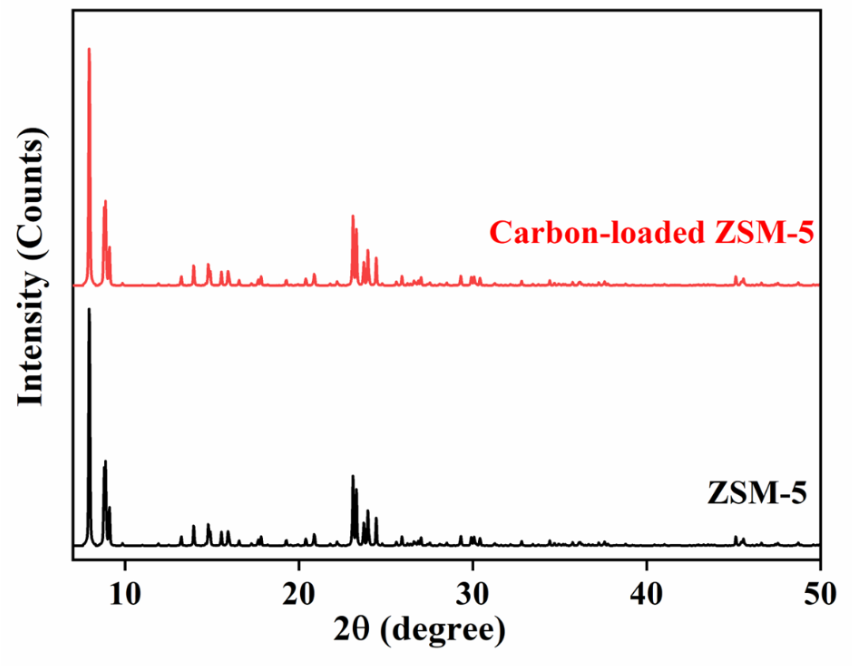}
  \caption{}
  \label{fig3_chim}
\end{subfigure}\hfill
\begin{subfigure}{0.48\textwidth}
  \centering
  \includegraphics[height=5cm, width=\linewidth, keepaspectratio]{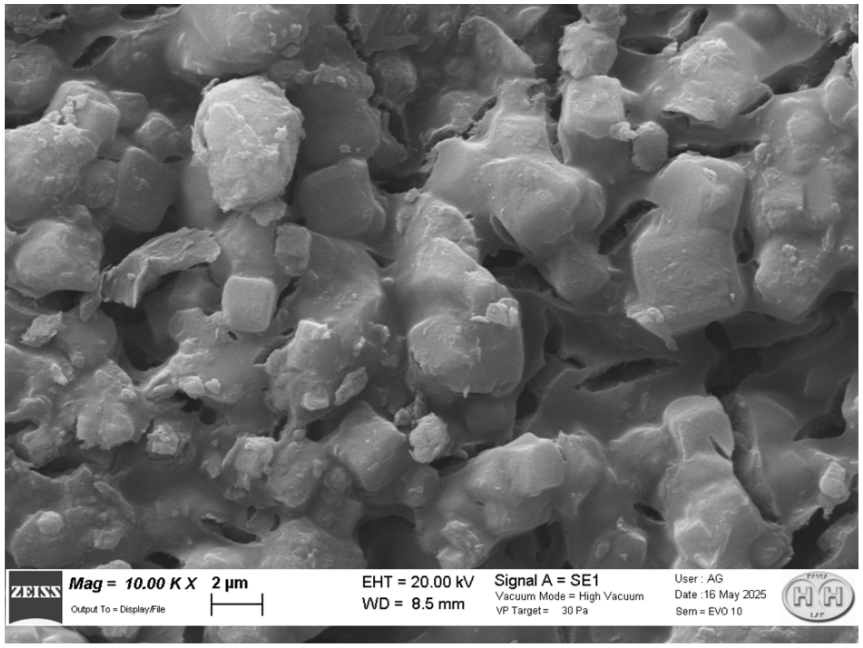}
  \caption{}
  \label{fig4_chim}
\end{subfigure}
\begin{subfigure}{0.48\textwidth}
  \centering
  \includegraphics[height=5cm, width=\linewidth, keepaspectratio]{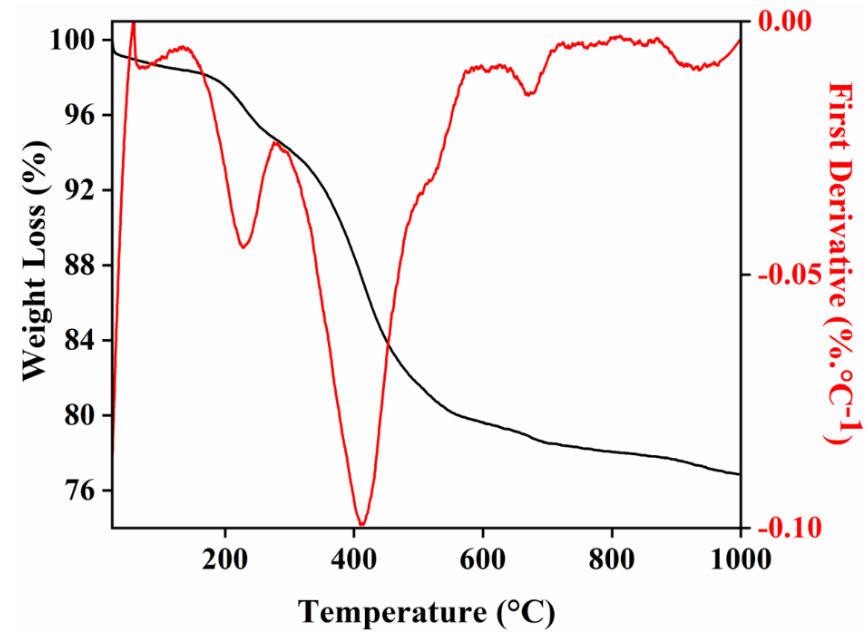}
  \caption{}
  \label{fig5_chim}
\end{subfigure}
\caption{(a) Comparison of PXRD patterns of commercial Z5 and carbon-loaded Z5 zeolite. (b) SEM image of carbon-loaded Z5 zeolite at 10000x magnification. (c) TGA of carbon-loaded Z5 zeolite in a static air atmosphere.}
\label{fig:carbonLoaded_Z5}
\end{figure}

To further enhance methane adsorption, zeolite can be combined with carbon material, as indicated by literature review. 
Therefore, a resulting 16,3\% carbon loaded Z5 zeolite was synthesized using kraft lignin as the carbon precursor~\cite{Z1,Z2}, adapting a procedure from a previous work~\cite{Z3,Z4}. In particular,   \SI{4}{g} of Z5 molecular sieves and \SI{1}{g} of low-sulfur content kraft lignin (Sigma-Aldrich, USA) were ground together until homogeneous, to prepare nominally \SI{5}{g} of 20\% carbon-loaded Z5 sample. Kraft lignin was selected as carbon precursor as it is an organic polymer extremely rich in carbon: adding it in the mixture assures that, after carbonization, solid carbon will be deposited inside and above the zeolite structure. Moreover the usage of low-sulfur lignin provides a higher degree of purity of the active site, which will then result in a more stable carbon residual. The resulting mixture was then dispersed in \SI{50}{\milli\liter} of distilled water and sonicated for 10 minutes to ensure uniform blending. The water was subsequently removed using a rotary evaporator, and the resulting powder was pyrolysed under vacuum at \SI{550}{\celsius}. The resulting material was characterised via PXRD to assess phase purity and crystallinity, and by Thermogravimetric Analysis (TGA) in a static air atmosphere to quantify the final carbon yield. The PXRD pattern of the carbon-loaded zeolite (Fig.~\ref{fig3_chim}), demonstrates that neither the carbon loading nor the pyrolysis at \SI{550}{\celsius} adversely affected the phase purity or crystallinity of the Z5 framework~\cite{Z5}. This structural preservation was corroborated by SEM imaging (Fig.~\ref{fig4_chim}), which shows the Z5 surface coated with carbon layers without obstructing the pore entrances. The actual carbon-loaded on Z5 zeolite percentage was determined via TGA, as carbon oxidises at elevated temperatures and evolves as carbon dioxide~\cite{Z6}. The TGA curve and its first derivative (Fig.~\ref{fig5_chim}) exhibit a major weight loss of 16.3\% near \SI{400}{\celsius}, corresponding to the final carbon content. This microporous carbon layer provides an extended surface area that can favorably interact with non-polar methane, thereby enhancing overall gas adsorption within the zeolite pores~\cite{Z7}.  
 
Literature also highlights other materials with high potential for methane adsorption at higher concentrations, notably the gallium/germanium zeolite UCSB-9 and the Zeolitic Imidazole Framework (MOF) TUT-100 (Taiyuan University of Technology-100). Because the synthesis of these advanced materials, as well as the one of carbon-loaded zeolites, faces significant scalability challenges, only the commercial Z5 zeolite was utilized for the primary methane adsorption experiments in this study. Nevertheless, UCSB-9 and TUT-100 are studied here primarily to provide prospective researchers with insights into diverse, high-performance materials for methane separation.     

\begin{figure}[htbp]
\centering
\begin{subfigure}{0.50\textwidth}
  \centering
  \includegraphics[height=5cm, width=\linewidth, keepaspectratio]{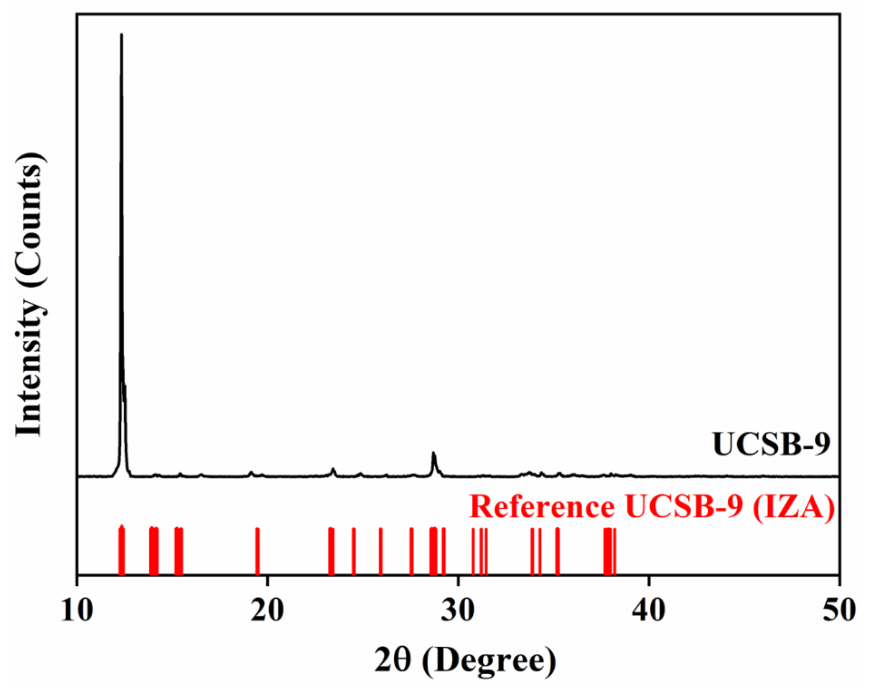}
  \caption{}
  \label{fig6_chim}
\end{subfigure}\hfill
\begin{subfigure}{0.48\textwidth}
  \centering
  \includegraphics[height=5cm, width=\linewidth, keepaspectratio]{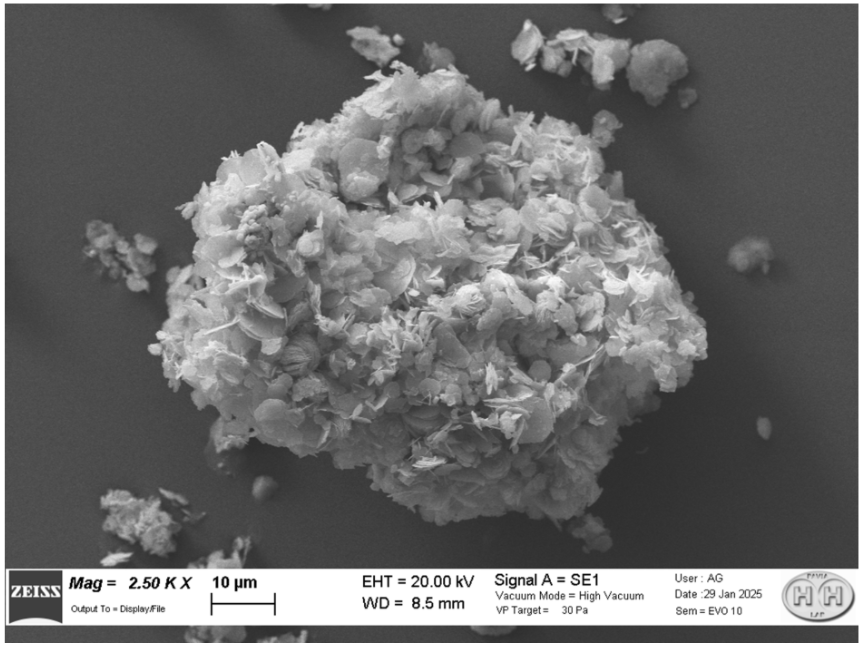}
  \caption{}
  \label{fig7_chim}
\end{subfigure}
\caption{(a) Comparison of PXRD patterns of prepared UCSB-9 zeolite and reference patterns obtained from the IZA. (b) SEM image of  UCSB-9 zeolite at 25000x magnification.}
\label{fig:uscb9}
\end{figure}

UCSB-9 is the first 3-ring gallogermanate zeolite of the SBN topology type; it contains two 8-ring pore openings along [001] and [101] axes and one 9-ring pore opening along the [010] axis~\cite{Z8}. In the context of this study, it was hydrothermally synthesized according to establish protocols~\cite{Z9} and its phase purity and crystallinity were confirmed by comparing the acquired PXRD data with IZA reference patterns (Fig.~\ref{fig6_chim})~\cite{Z10}. 
SEM imaging of the synthesized UCSB-9 (Fig.~\ref{fig7_chim}) revealed hierarchically assembled microspherical flakes, providing a high surface area ideal for methane adsorption.

\begin{figure}[htbp]
\centering
\begin{subfigure}{0.50\textwidth}
  \centering
  \includegraphics[height=5cm, width=\linewidth, keepaspectratio]{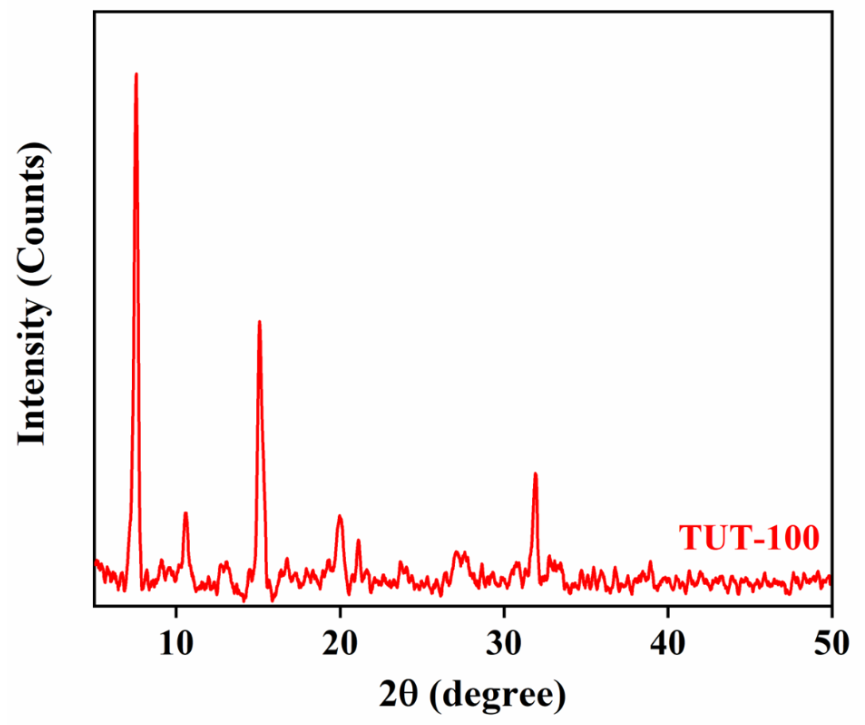}
  \caption{}
  \label{fig8_chim}
\end{subfigure}\hfill
\begin{subfigure}{0.48\textwidth}
  \centering
  \includegraphics[height=5cm, width=\linewidth, keepaspectratio]{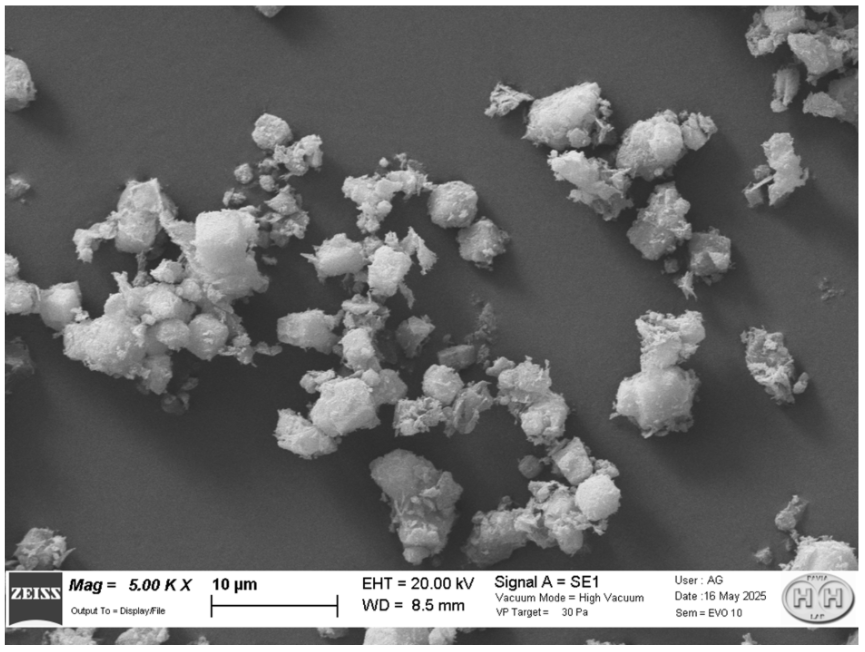}
  \caption{}
  \label{fig9_chim}
\end{subfigure}
\caption{(a) PXRD of TUT-100 ZIF. (b) SEM image of  TUT-100 ZIF at 5000x magnification.}
\label{fig:tut}
\end{figure}

Similarly, TUT-100, a cobalt-based 4,5-dichloroimidazole Zeolitic Imidazolate Framework (ZIF) with a sodalite cage size of \SI{0.8}{nm}, was selected for its reported high methane uptake (\SI{45.29}{cm^3/cm^3}), high \ce{CH4}/\ce{N2} selectivity (6.3 at \SI{298}{K} and \SI{1}{bar}), and hydrophobicity  ($<$\SI{6.3}{cm^3/g} \ce{H2O}  adsorption at \SI{298}{\kelvin}). TUT-100 was prepared following the hydrothermal synthesis procedure presented in Ref.~\cite{Z11} and its structural integrity was validated by PXRD (Fig.~\ref{fig8_chim}) and SEM imaging (Fig~\ref{fig9_chim}), both of which align well with the literature~\cite{Z11}. 

\section{Technology validation with laboratory scale prototype}\label{sec:lab_proto}

The primary objectives of the technology validation phase are to assess the feasibility of methane adsorption in a livestock housing environment and to determine the adsorption capacity of \ce{CH4} under various operating conditions. These aspects are critical for the design of the final prototype intended for real-world validation. To achieve this, a dedicated laboratory setup was engineered to allow independent variation and precise control of key parameters, enabling an accurate assessment of their individual impacts on adsorption performance.

The initial phase of this study aimed to demonstrate the inherent capability of zeolites to adsorb methane.
Subsequently, the focus shifted to quantifying the \ce{CH4} uptake across different commercial zeolite types to identify the optimal candidate for practical deployment. Furthermore, the study evaluated  the regeneration of the adsorbent materials, specifically analyzing process duration and efficiency to minimize both power consumption and system downtime. Finally, environmental and operational parameters --- namely humidity, methane concentration, and system pressure --- were systematically varied to determine their influence on overall adsorption capacity.

\subsection{Laboratory-scale prototype design}

\begin{figure}[!h]
    \centering
    \includegraphics[width=1\linewidth]{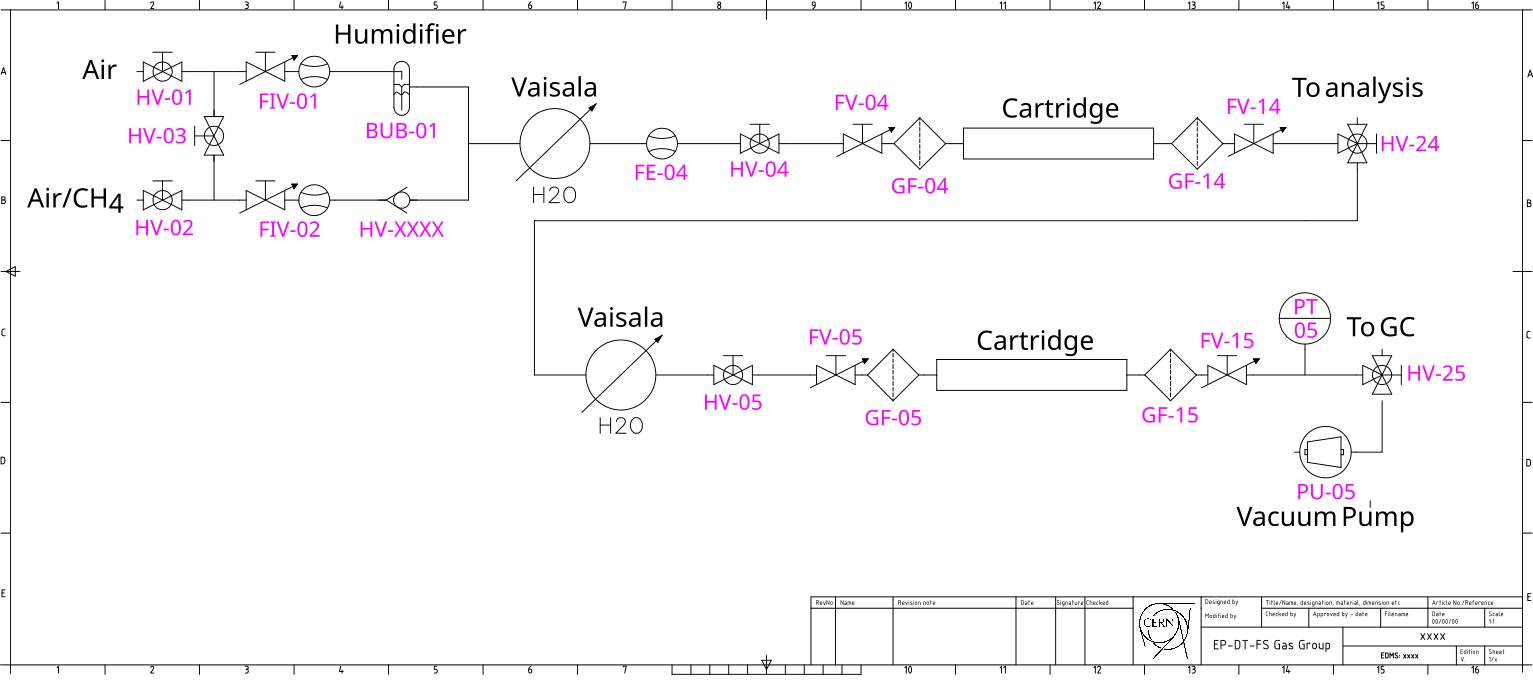}
    \caption{Schematic representation of the early prototype developed in the lab.}
    \label{fig:schema_lab_1}
\end{figure}

A custom laboratory-scale prototype was designed to conduct the adsorption experiments, as depicted in Fig.~\ref{fig:schema_lab_1}. The system can be conceptually divided into three main sections: gas supply and conditioning, adsorption stage, and analysis and regeneration line.
The gas supply section comprises two distinct lines: the primary carrier gas line utilizes compressed air to modulate both the total flow rate and the relative humidity of the final mixture. The secondary line acts as the methane source, supplying a calibrated \ce{CH4}-in-air mixture at a concentration of \SI{4509}{ppm}.
The flow rate of each line is regulated by manual valves (HV-01, HV-02) and measured by dedicated rotameters (FIV-01, FIV-02) with a measuring range of \SI{3}{\liter/h}, allowing fine control over the blending ratio. To simulate varying environmental moisture levels, the compressed air stream is routed through a water bubbler (BUB-01).
A check valve (HV-xxxx) on the methane line prevents potential backflow of moisture. The two streams are then merged via standard T-junctions.
Downstream of the mixing point, the gas conditioning is monitored using a VAISALA DMT143 dew point sensor~\cite{vaisala} to measure humidity, followed by an OMRON D6F-P0010A2 mass flow meter (FE-04). An isolation valve (HV-04) controls the injection of the gas mixture into the adsorption stage. This stage accommodates the zeolite cartridges, which are isolated by needle valves (FV-04, FV-14, FV-05, FV-15) and protected by inline particulate filters (GF-04, GF-14, GF-05, GF-15) to prevent the dispersion of zeolite dust. A three-way directional valve (HV-24) is positioned after the first cartridge, allowing the effluent gas to be routed either sequentially into the second cartridge or directly to the analysis line. A secondary humidity sensor~\cite{vaisala} is installed after the valve before the second cartridge. At the system's terminus, a pressure transducer (PT05) monitors the operational pressure, while a final three-way valve (HV-25) directs the flow either to the analysis equipment (to gas chromatograph - GC) or to a vacuum pump, which is employed for the desorption and regeneration of saturated cartridges.

\begin{figure}[htbp]
\centering
\begin{subfigure}{0.5\textwidth}
  \centering
  \includegraphics[width=0.96\linewidth, keepaspectratio]{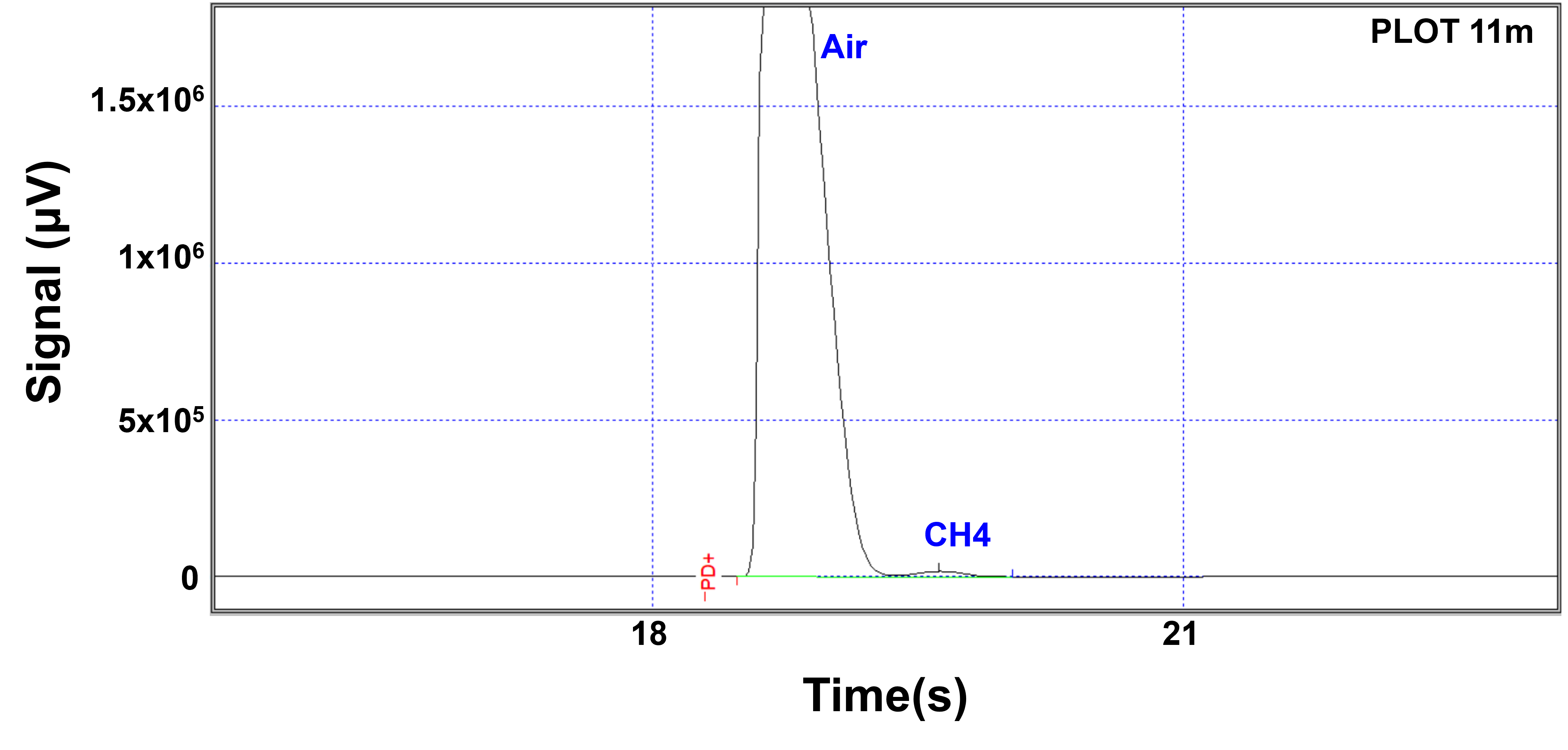}
  \caption{}
\end{subfigure}
\begin{subfigure}{0.48\textwidth}
  \centering
  \includegraphics[width=\linewidth, keepaspectratio]{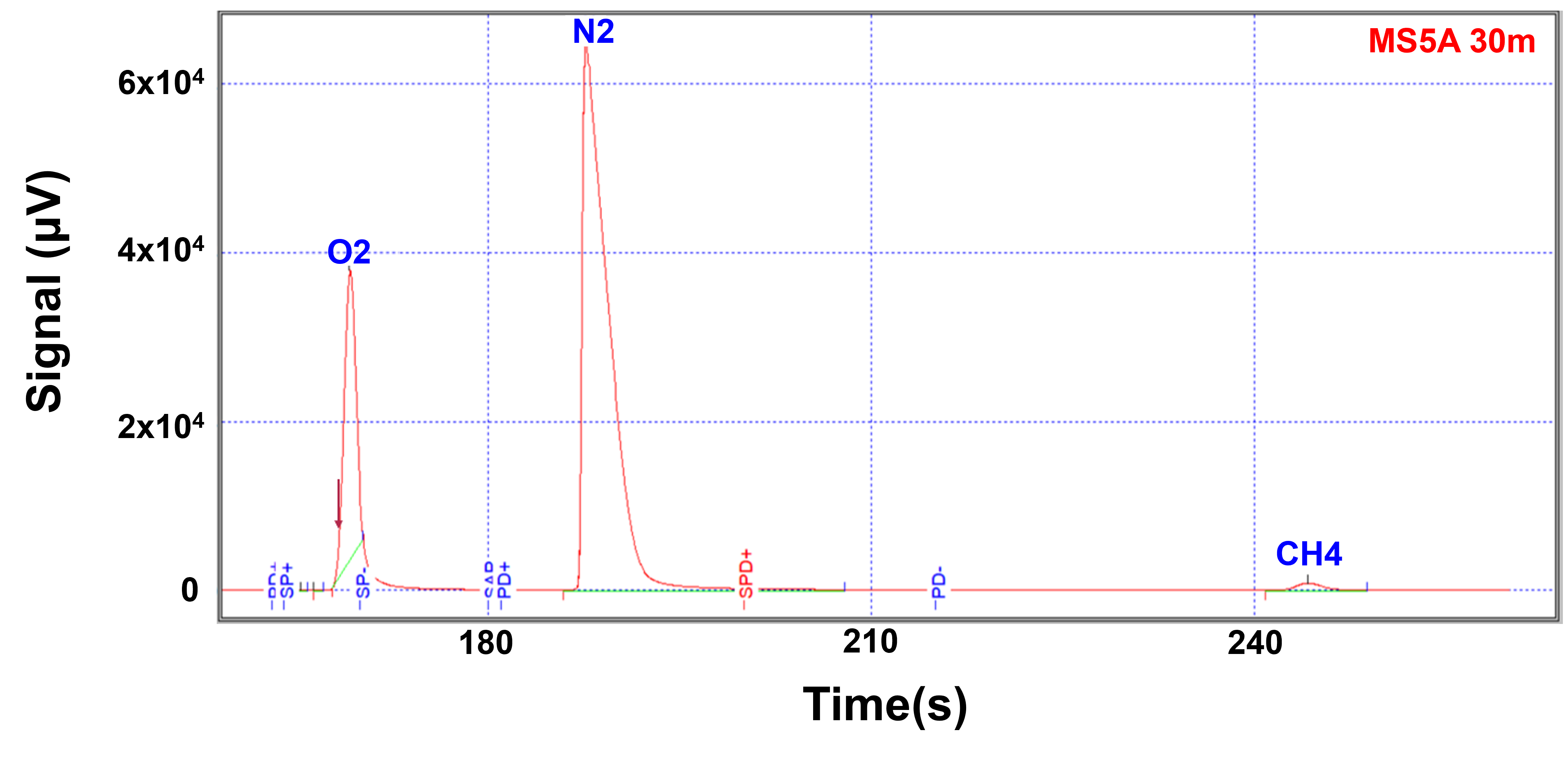}
  \caption{}
\end{subfigure}
\caption{Comparison of typical chromatograms: (a) PPU and (b) MS columns. In the PPU, the \ce{CH4} peak overlaps with the dominant air signal due to similar retention times. Conversely, the MS column achieves baseline separation of \ce{CH4} from \ce{N2} and \ce{O2}.}
\label{fig:PPU_MS_cromatogrammi_esempio}
\end{figure}

Gas composition analysis is performed using a GC~\cite{gc}.
The GC separates \ce{CH4} from the bulk mixture using either a Porous Polymer Urea (PPU) of type Porous Layer Open Tubular (PLOT), often referred as CP-PoraPLOT U, or a Molecular Sieve (MS) column, outputting an electrical signal proportional to the methane concentration. Representative chromatograms obtained with these two columns are shown in Fig.~\ref{fig:PPU_MS_cromatogrammi_esempio}.

While both are specialized columns for separating permanent gases and low-boiling hydrocarbons at or above ambient temperatures, they rely on distinct separation mechanisms. PLOT columns feature a $5-\SI{50}{\micro\meter}$ layer of porous particles adhered to the inner wall of a capillary (typically 0.32 or \SI{0.53}{mm} inner diameter). This architecture is highly effective for detecting volatile compounds, eliminates the need for cryogenic cooling for permanent gases, and exhibits high selectivity for isomers. This type of columns are primarily employed for refinery gases, permanent gases (\ce{H2}, \ce{O2}, \ce{N2}, \ce{CH4}, \ce{CO}), and light hydrocarbons. 

In contrast, MS columns are specifically designed to resolve permanent gases such as argon, oxygen, nitrogen, and carbon monoxide. Commonly utilized 5A (\SI{0.5}{nm} pore) and 13X (\SI{1}{nm} pore) particles separate species based on molecular size. The primary limitation of MS columns is their high sensitivity to water and carbon dioxide, which can lead to the progressive deactivation of their adsorption capacity.

\subsection{Comparison of commercial zeolites performance}\label{confronto_zeoliti}

The initial series of measurements aimed to determine the most suitable zeolites for humidity and \ce{CH4} adsorption, selecting from the following commercially available materials previously utilized at CERN:
\begin{itemize}
    \item MS with \SI{3}{\angstrom} pore size, designated as Z3;
    \item MS with \SI{4}{\angstrom} pore size (Z4);
    \item MS with \SI{5}{\angstrom} pore size (Z5);
    \item 13X MS with \SI{10}{\angstrom} pore size (Z10).
\end{itemize}

The first three (Z3, Z4, Z5) have the same crystal structure, called Linde Type A (LTA)~\cite{ref:lta}, with identical basic structure but with different charge-compensating cation, which influences the different pore sizes. 13X instead belongs to a completely different family of materials, called Faujasite (FAU)~\cite{ref:faujasite}. The ``X'' indicates a low Si/Al ratio in its composition, which makes this zeolite extremely polar.

\begin{figure}[!h]
    \centering
    \includegraphics[width=1\linewidth]{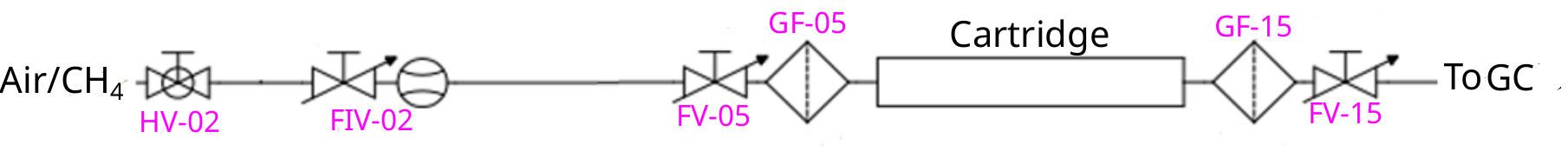}
    \caption{Schematic of the setup used to evaluate the adsorption capacity of the various zeolites~\cite{Tesi_Francesco}.}
    \label{fig:setup_primi_test}
\end{figure}

\begin{figure}[htbp]
\centering
\begin{subfigure}{0.48\textwidth}
  \centering
  \includegraphics[width=\linewidth]{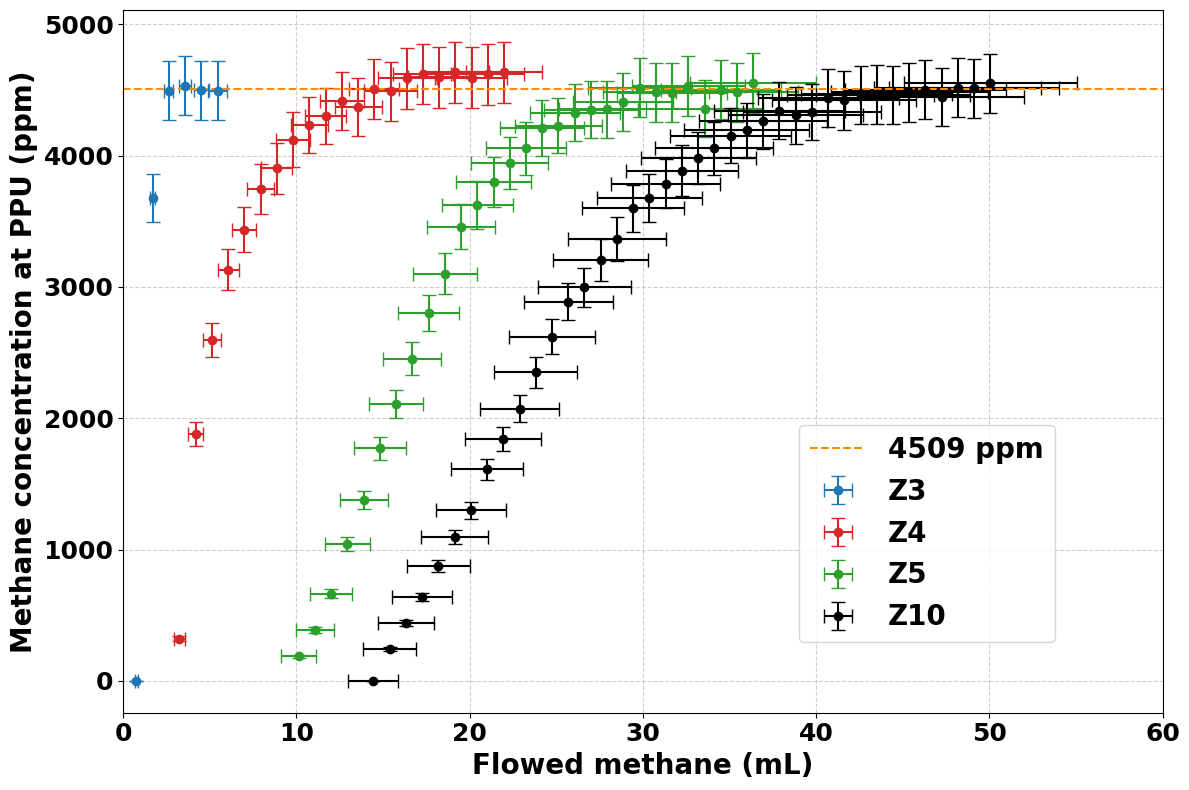}
  \caption{}
  \label{fig:saturation_curves_PPU}
\end{subfigure}\hfill
\begin{subfigure}{0.48\textwidth}
  \centering
  \includegraphics[width=\linewidth]{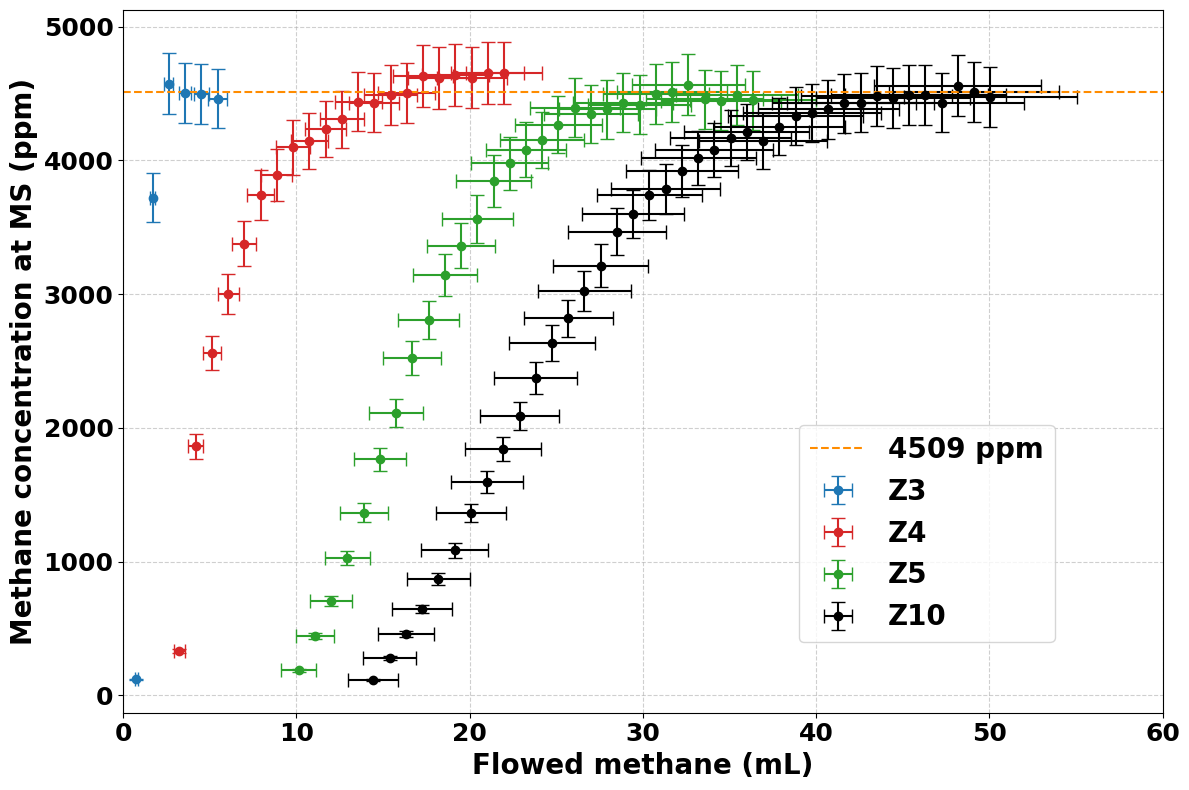}
  \caption{}
  \label{fig:saturation_curves_MS}
\end{subfigure}
\caption{Methane concentration sampled at the cartridge outlet for various zeolite samples, analyzed using the GC PPU (a) and MS (b) columns. The orange dashed horizontal line represents the nominal \ce{CH4} concentration at the input (\SI{4509}{ppm}).}
\label{fig:saturation_curves_diverseZeoliti}
\end{figure}

The experimental setup is depicted in Fig.~\ref{fig:setup_primi_test}: each zeolite sample (approximately \SI{270}{g}) was housed in a metal test cartridge and flushed with a gas mixture containing \SI{4509}{ppm} of methane. The effluent gas was subsequently analyzed via GC. As illustrated in Fig.~\ref{fig:saturation_curves_diverseZeoliti}, in each run, an initial phase is observed during which no methane is detected at the cartridge outlet, indicating complete adsorption by the zeolite. The breakthrough (BK) point is defined as the last instant prior to \ce{CH4} detection at the outlet; after the BK, the concentration measured at the outlet steadily increases, until it equilibrates with the supply \ce{CH4} concentration. 

If the GC run is composed of $n$ samplings up to saturation, each of duration $\Delta t_i$ (with $i=0,1,...,n$), and $i = \text{BK}$ ($\text{BK} < n$) is the breakthrough instant, the volume of adsorbed methane can be determined using Eq.~\eqref{eq:V_metano_breakthrough}, for the volume adsorbed up to the BK point, and Eq.~\eqref{eq:V_metano_totale}, for the total volume of methane adsorbed in the entire run:

\begin{equation}
    V_{CH_4}^{BK} =  \sum_{i=0}^{BK} \phi_i\Delta t_i p_i C_{CH_4 }^{SAT} \ ,
    \label{eq:V_metano_breakthrough}
\end{equation}
\begin{equation}
    \begin{split}
        V_{\ce{CH4}}^{Tot} &= V_{\ce{CH4}}^{Flushed} - \int \text{GC saturation curve} \\ 
        &= \sum_{i=0}^{n} \phi_i \Delta t_i p_i C_{\ce{CH4}}^{SAT} - \sum_{i=0}^{n} \phi_i \Delta t_i p_i \text{GC}_i,
    \end{split}
    \label{eq:V_metano_totale}
\end{equation}
where $\phi_i$ is the gas flow, p$_i$ is the gas pressure and GC$_i$ is the GC measurement in the relevant i-th sampling interval. C$_{\ce{CH4}}^{SAT}$ is the nominal \ce{CH4} concentration in the gas. In this first set of tests the pressure of the gas is maintained constant at 1 bar, while C$_{\ce{CH4}}^{SAT}$ = 4509 ppm.

Initial tests revealed that Z3 saturated almost instantaneously; consequently, it was excluded from further evaluation. Table~\ref{tab:BT_diverseCartucce} summarizes the breakthrough volumes for the remaining zeolites.
Based on these findings, Z4 exhibited the lowest adsorption capacity, whereas Z5 and Z10 demonstrated comparable performance.

\begin{table}[]
\centering
\caption{\ce{CH4} volume adsorbed by the respective zeolites prior to breakthrough. }
\begin{tabular}{ll}
\hline
Zeolite & \begin{tabular}[c]{@{}l@{}}Breakthrough \\volume (mL)\end{tabular} \\ \hline
Z3  & $1.13 \pm 0.23$\\
Z4  & $3.13 \pm 0.60$\\
Z5  & $10.52 \pm 1.88$\\
Z10 & $12.10 \pm 1.72$\\ 
\hline
\end{tabular}
\label{tab:BT_diverseCartucce}
\end{table}
\subsection{Zeolites regeneration}
Subsequent testing focused on the regeneration of zeolites following methane adsorption. For this process, Z5 and Z10 were evaluated, as they demonstrated the most promising performance. The methods available for cartridge regeneration were: 
\begin{itemize}
    \item Thermal Swing Adsorption (TSA);
    \item Vacuum Swing Adsorption (VSA).
\end{itemize}
The objective of these tests was to determine whether the adsorption capacity of the zeolites could be fully restored using VSA, a method that offers lower energy consumption compared to TSA. Therefore, during the experimental procedure, the cartridges were saturated with methane and subsequently evacuated using a vacuum pump at a pressure of $-\SI{970}{mbar}$. This cycle was repeated several times. During each iteration, the volume of methane required to re-saturate the material was recorded. 

\begin{figure}[h]
    \centering
    \includegraphics[width=0.7\linewidth]{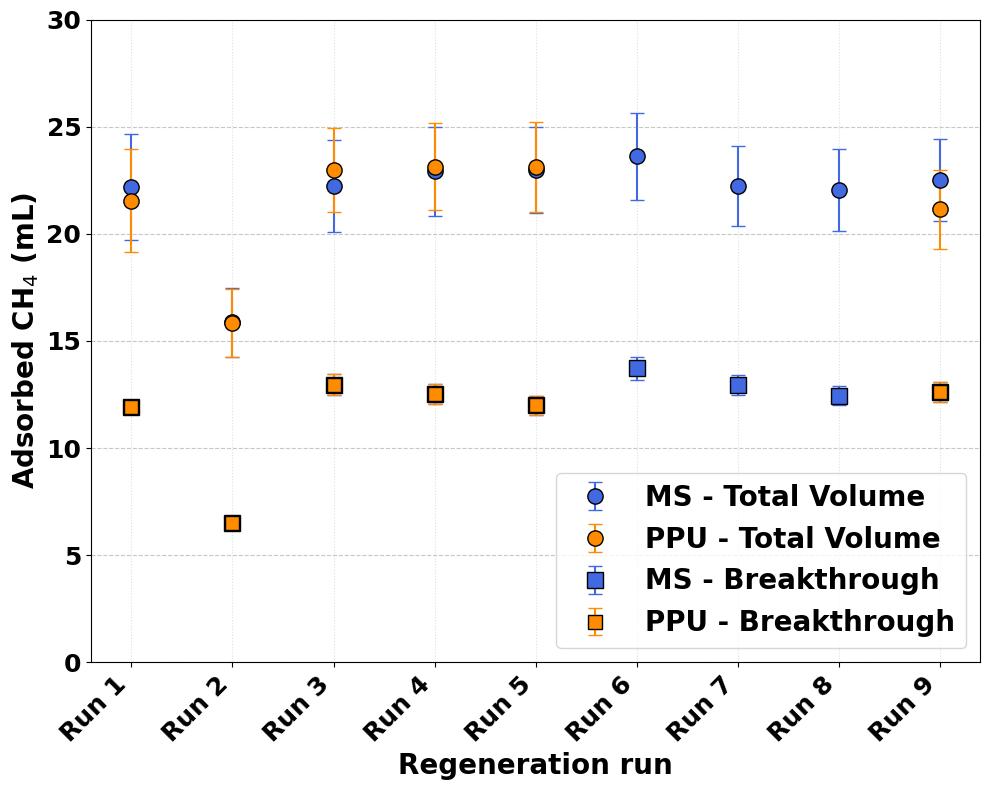}
    \caption{Volume of methane adsorbed by Z5 following VSA regeneration cycles of 40 minutes each. Run 1 represents the adsorption of Z5 after TSA. Only MS column results are available for run 6, 7 and 8.}
    \label{fig:adsorbed_novembre_secco}
\end{figure}

Results are presented in Fig.~\ref{fig:adsorbed_novembre_secco}: run 1 was performed after a TSA regeneration and it is taken as reference. The following runs, performed after VSA regeneration, demonstrate that Z5 maintained an adsorption capacity consistent with that achieved with TSA. Furthermore, the adsorption capacity remained stable over time, even after multiple VSA cycles. Run 2 deviates from the average due to a slight contamination of the gas lines with \ce{CH4} from the previous run, thus a lower result was expected.  

Conversely, the adsorption capacity of Z10 was observed to decrease by approximately one-third over multiple VSA cycles. This degradation suggests that VSA is not entirely effective for regenerating Z10. Consequently, Z5 was selected as the best candidate for the subsequent stages of the project. 

\subsection{Humidity adsorption}
Due to its low methane adsorption capacity and strong affinity for moisture (with a water adsorption capacity of \SI{140}{g/kg}), Z3 was chosen as the optimal candidate for capturing humidity. These characteristics are highly advantageous for the prototype, as they allow for selective humidity sequestration within a dedicated cartridge, permitting methane to pass through and be captured in a secondary Z5 stage.

\begin{figure}[h]
    \centering
    \includegraphics[width=\linewidth]{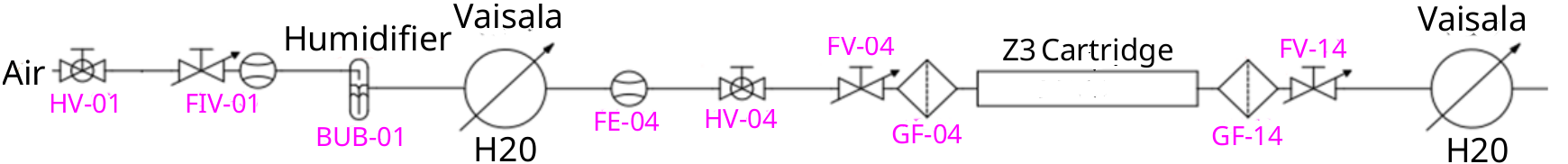}
    \caption{Experimental setup for evaluating  humidity adsorption by Z3.}
    \label{fig:setup_Z3_humidity_testing}
\end{figure}

The adsorption efficiency of Z3 was evaluated using the setup depicted in Fig.~\ref{fig:setup_Z3_humidity_testing}, where humidity was introduced into the gas mixture via a water bubbler. Humidity levels were monitored upstream (\SI{22600}{ppm})  and  downstream (\SI{20}{ppm}) of the cartridge, demonstrating the nearly complete removal of humidity from the gas stream upon passing through the Z3.  

\subsection{\ce{CH4} adsorption capabilities at different methane concentrations}

The volume of adsorbed methane is dependent on its partial pressure within the mixture, and thus on its concentration. Evaluating adsorbed methane across varying concentrations is critical for predicting  adsorption capacity in a real-world scenario, particularly since replicating the expected concentrations in the barn during testing was unfeasible. The primary goal of these tests was to quantify the methane captured by Z5 at various concentrations  and to extrapolate these findings to the parts-per-million (tens or hundreds of ppm) range anticipated in the barn. Direct measurement in this low-concentration regime was precluded by the sensitivity limits of the GC and the physical constraints of diluting the methane mixture using rotameters.

\begin{figure}[h]
    \centering
    \includegraphics[width=\linewidth]{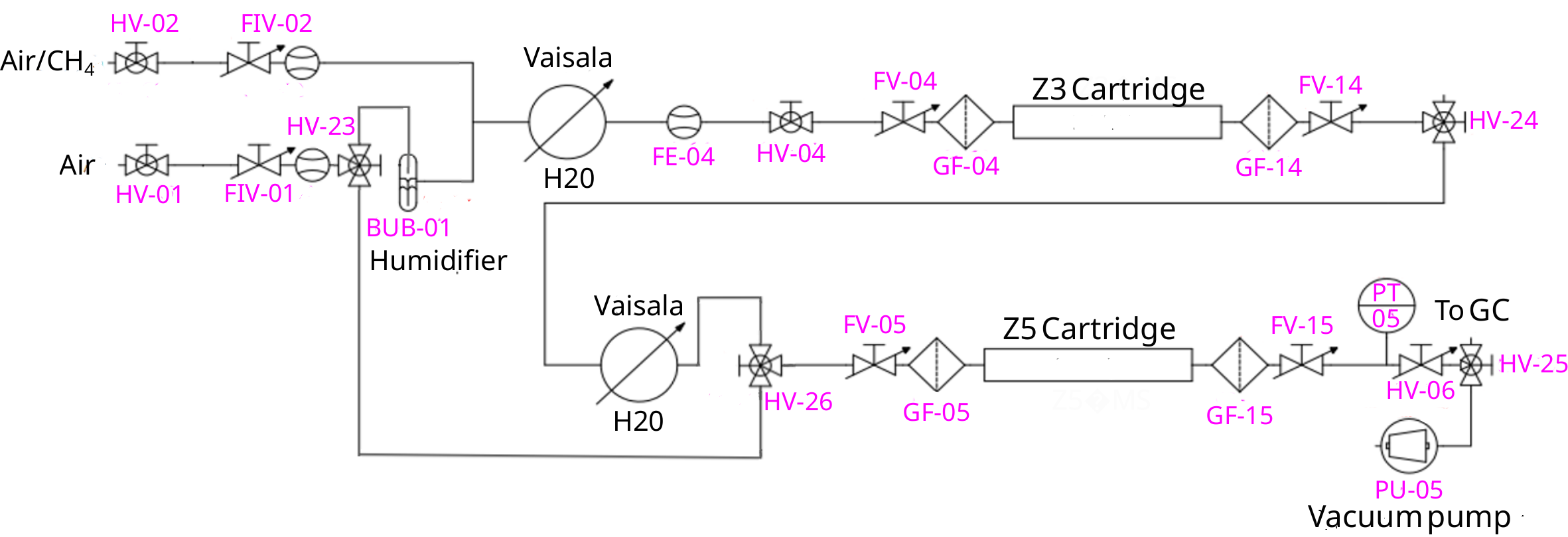}
    \caption{Experimental setup employed for the measurement of the baseline \ce{CH4} concentration of \SI{4509}{ppm}, used as reference. Variable \ce{CH4} concentrations are then achieved by diluting the source gas with supplementary air. The setup used for the measurement of diluted gases is identical to the one depicted, with the exception of the humidifier and Z3 cartridge, which were removed. A bypass line connecting the air cylinder to the Z5 cartridge inlet is utilized to refill the cartridge with air following the vacuum regeneration.}
    \label{fig:Schema_diluizione_con_bypass}
\end{figure}

The experimental protocol, performed with the setup in Fig.~\ref{fig:Schema_diluizione_con_bypass}, was established as follows: initially, a reference curve, with \SI{4509}{ppm}, was recorded as a baseline using a freshly regenerated Z5 cartridge. This baseline measurement was performed multiple times, every time with a freshly regenerated Z5, to ensure data consistency. 
Subsequently, the methane source was diluted with air from the compressed air cylinder (consisting strictly of atmospheric ratios of nitrogen and oxygen).
The input concentration was dictated by the flux ratio of the rotameters on the respective  input lines; for instance, achieving a 50\% dilution required equalizing the flux from both the methane source and the compressed air cylinder. System geometry limited the range of viable concentrations, as disproportionately opening one rotameter caused the opposing line to close. 

In total, six distinct concentrations were evaluated, ranging from  \SI{4509}{ppm} down to $\simeq \SI{900}{ppm}$. 

\begin{table}[htbp]
\setlength{\tabcolsep}{3pt} 
\centering
\caption{Adsorption capacity evaluated across various methane concentrations (C). Data points represent the average of three independent trials, with the exception of the standard reference (six trials) and the \SI{3615}{ppm} data point (a single measurement). The errors on each measurement were obtained from the error propagation applied to Eq.~\eqref{eq:V_metano_breakthrough}~and~\eqref{eq:V_metano_totale}. The error on the average value is then the standard deviation on the mean. The tests were performed with \SI{270}{g} of Z5.}
\begin{tabular}{ccccc}

\hline
C [ppm]& BK PPU [mL] & BK MS [mL] & Tot Vol PPU [mL] & Tot Vol MS [mL] \\
\hline
$4509 \pm 0$ & $15.72 \pm 0.08$ & $16.85 \pm 0.08$ & $23.23 \pm 0.27$ & $23.46 \pm 0.28$ \\
$3615 \pm 25$ & $14.35 \pm 0.16$ & $14.35 \pm 0.16$ & $19.41 \pm 0.51$ & $19.25 \pm 0.51$ \\
$2531 \pm 2$ & $9.97 \pm 0.06$ & $10.38 \pm 0.07$ & $13.42 \pm 0.22$ & $13.47 \pm 0.21$ \\
$1703 \pm 6$ & $6.80 \pm 0.04$ & $7.23 \pm 0.05$ & $8.82 \pm 0.14$ & $8.84 \pm 0.14$ \\
$1284 \pm 6$ & $5.00 \pm 0.03$ & $5.22 \pm 0.03$ & $6.48 \pm 0.11$ & $6.56 \pm 0.11$ \\
$899 \pm 2$ &  $4.02 \pm 0.03$ & $4.09 \pm 0.03$ & $4.91 \pm 0.07$ & $4.90 \pm 0.07$ \\
\hline
\end{tabular}
\label{tab:misure_concentrazione}
\end{table}

Table~\ref{tab:misure_concentrazione} details the adsorption results corresponding to varying input methane concentrations. 
\begin{figure}[h]
    \centering
    \includegraphics[width=0.7\linewidth]{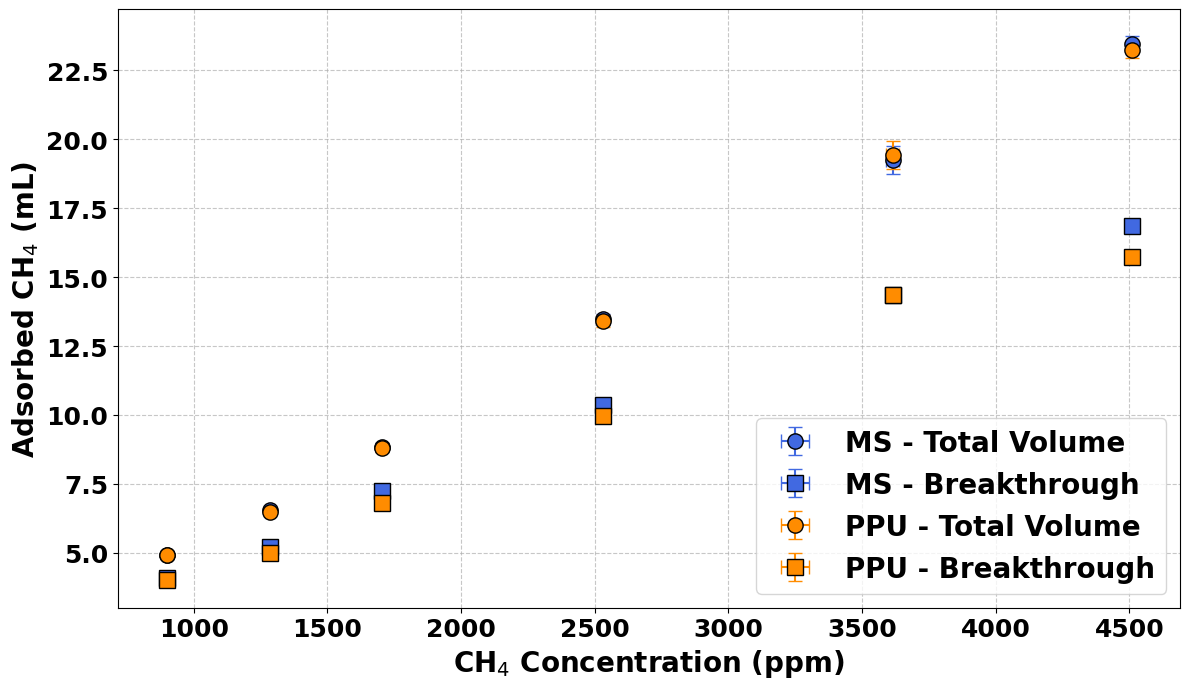}
    \caption{Adsorbed methane as a function of the input concentration. The error bars are contained in the markers.}
    \label{fig:concVSads}
\end{figure}
Furthermore, Fig.~\ref{fig:concVSads} illustrates the dependence of the adsorbed \ce{CH4} volume  on initial input  concentration.

\begin{figure}[htbp]
\centering
\begin{subfigure}{0.5\textwidth}
  \centering
  \includegraphics[width=\linewidth]{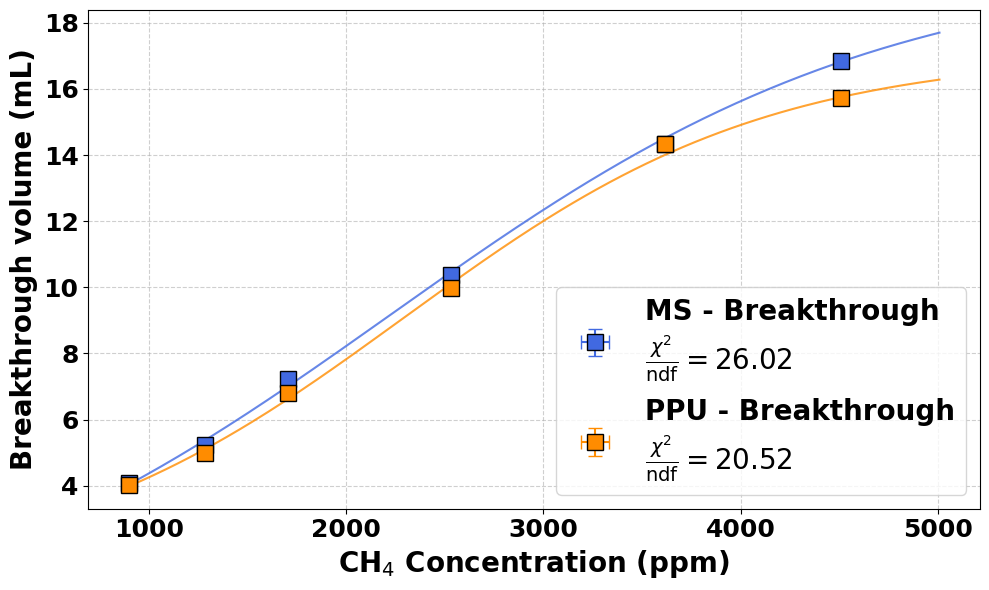}
  \caption{}
  \label{fig:Concentration_VS_BT_Volume}
\end{subfigure}\hfill
\begin{subfigure}{0.5\textwidth}
  \centering
  \includegraphics[width=\linewidth]{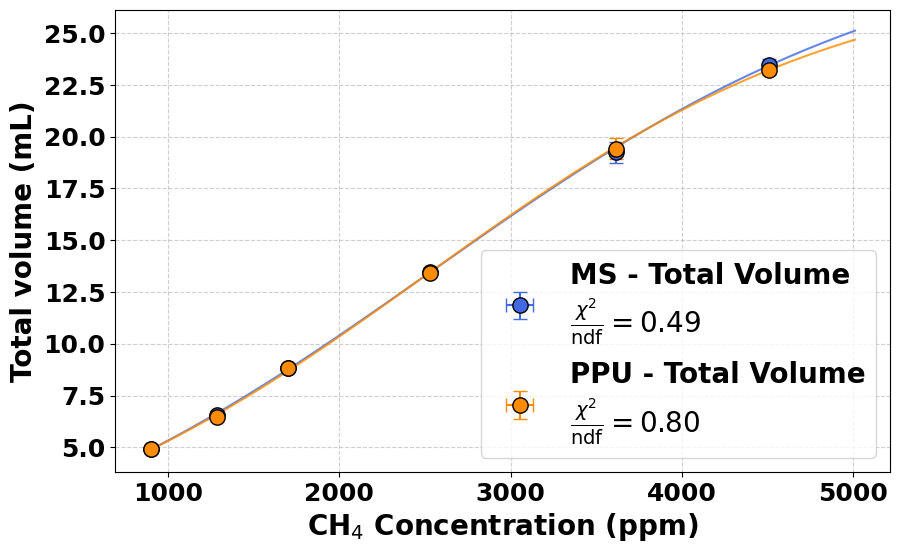}
  \caption{}
  \label{fig:Concentration_VS_Saturation_Volume}
\end{subfigure}
\caption{Volume of methane adsorbed by \SI{270}{g} of Z5 zeolite as a function of \ce{CH4} concentration in the gas mixture. (a) Volume of methane adsorbed up to the BK; (b) total volume adsorbed up to complete saturation. The error bars are included in the markers. }
\label{fig:Concentration_VS_Volume}
\end{figure}

The mathematical relationship between the volume of methane captured by the Z5 zeolites and its concentration in the gas mixture was modeled using the following sigmoid function:
\begin{equation}
    y = \frac{L}{1+e^{-k(x-x_0)}}+b.
    \label{eq:sigmoid}
\end{equation}
Fig.~\ref{fig:Concentration_VS_Volume} shows the results, while the fit parameters are detailed in Tab.~\ref{tab:sigmoid_parameters}.

\begin{table}[!h]
\centering
\caption{Parameters of the sigmoid fits in Eq.~\eqref{eq:sigmoid} applied to the data in Fig.~\ref{fig:Concentration_VS_Volume}.}
\resizebox{\textwidth}{!}{
\begin{tabular}{cccccc}
\hline
Data kind    & GC column & L (mL) & k (ppm$^{-1}$) & x$_0$ (ppm) & b (mL)\\ \hline
$V_{BK}$ & PPU       & $15.96\pm0.48$ & $(1.1\pm0.1)\times10^{-3}$ & $2295.38\pm25.19$ & $1.11\pm0.29$\\
$V_{BK}$ & MS        & $21.26\pm1.22$ & $(0.8\pm0.1)\times10^{-3}$ & $2224.65\pm45.34$ & $-1.46\pm0.73$ \\
$V_{Tot}$   & PPU       & $29.87\pm3.78$ & $(0.8\pm0.1)\times10^{-3}$ & $2511.13\pm90.41$ & $-1.59\pm1.86$ \\
$V_{Tot}$   & MS        & $33.20\pm5.32$ & $(0.7\pm0.1)\times10^{-3}$ & $2538.92\pm110.49$ & $-3.10\pm2.57$ \\ \hline
\end{tabular}%
}
\label{tab:sigmoid_parameters}
\end{table}

\begin{figure}[h]
    \centering
    \includegraphics[width=0.8\linewidth]{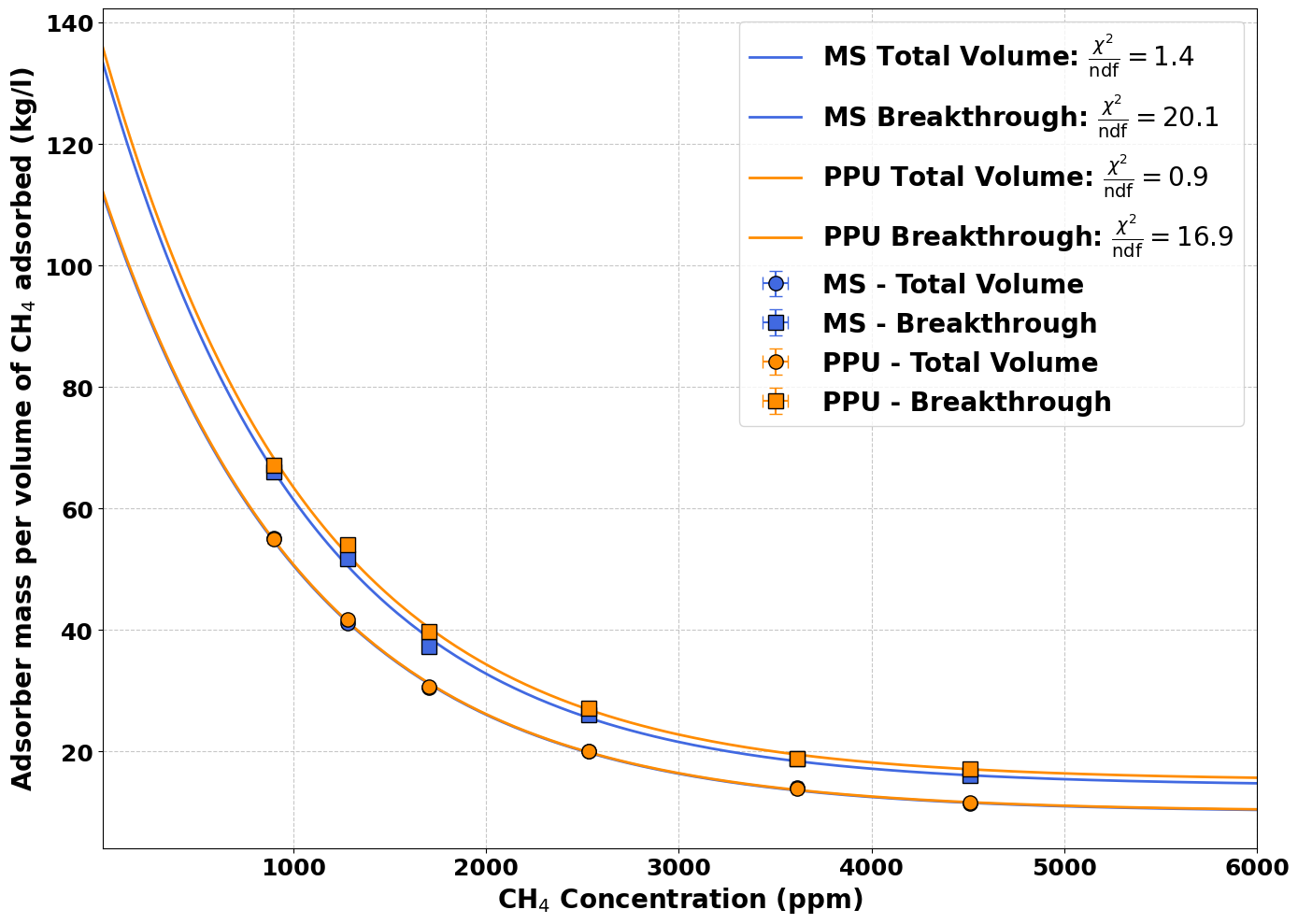}
    \caption{Mass of Z5 zeolite needed to adsorb one liter of methane as a function of \ce{CH4} concentration in the gas mixture. Estimates are derived from data obtained using both GC columns (PPU and MS) for methane adsorbed up to the breakthrough point or at  complete saturation of the Z5 cartridge. The fits are extrapolated down to \SI{10}{ppm} concentration.}
    \label{fig:mass_VS_concentration}
\end{figure}

Finally, in Fig.~\ref{fig:mass_VS_concentration} the estimate of Z5 mass required to adsorb one liter of methane as a function of the \ce{CH4} mixture concentration is presented. This projection was instrumental in establishing the required order of magnitude for the zeolite mass in the final prototype intended for installation in the barn.

\subsection{Study of adsorption as a function of pressure}

To quantify the influence of pressure on the methane adsorption capacity of Z5, several tests were conducted. The mass of Z5 used was consistent with previous tests, approximately of \SI{270}{g}. The absolute pressure was varied within a range of 1 to \SI{6}{bar}.

\begin{figure}[htbp]
\centering
\begin{subfigure}{0.5\textwidth}
  \centering
  \includegraphics[width=\linewidth]{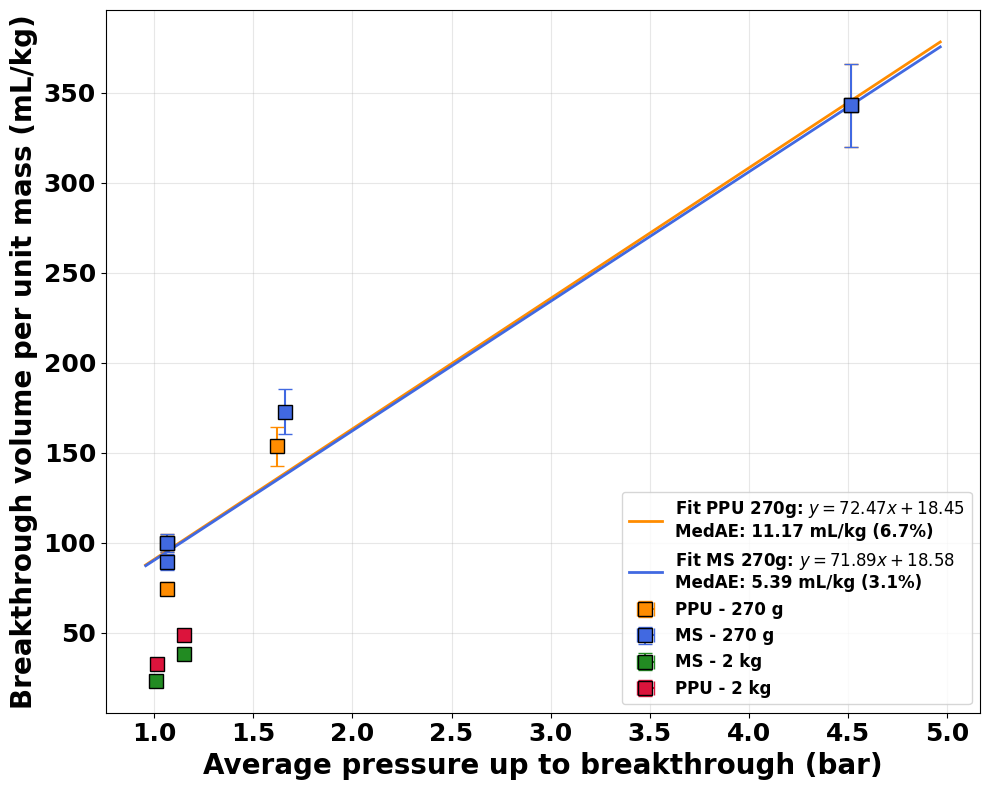}
  \caption{}
\end{subfigure}\hfill
\begin{subfigure}{0.5\textwidth}
  \centering
  \includegraphics[width=\linewidth]{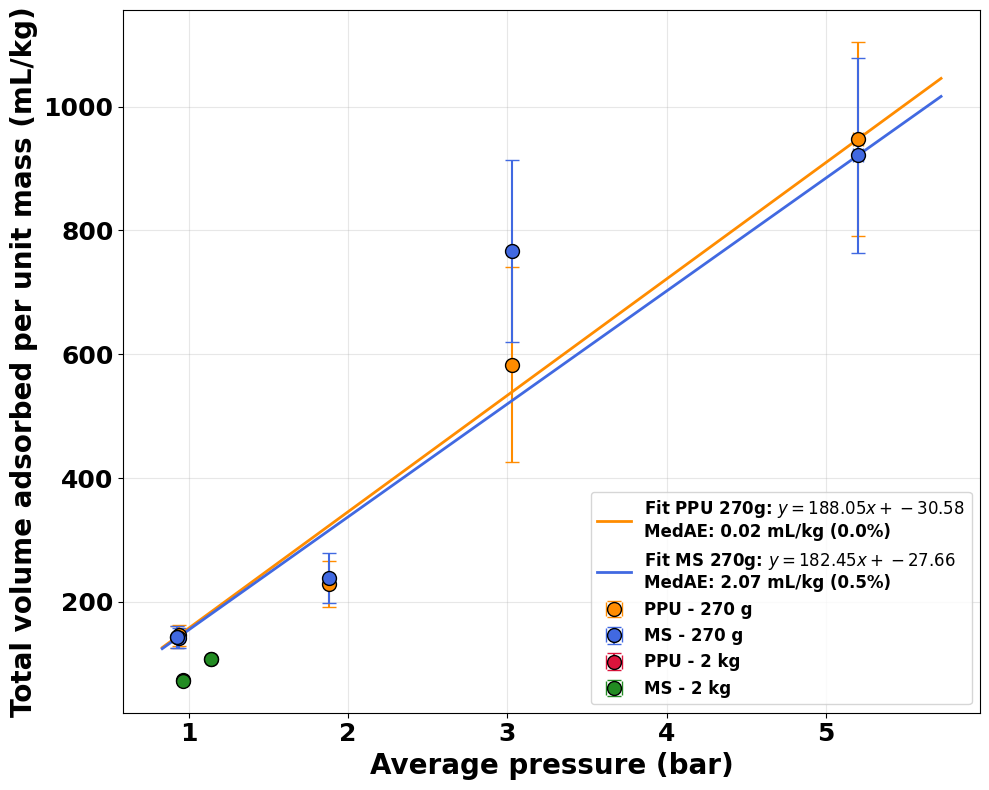}
  \caption{}
\end{subfigure}
\caption{Methane volume adsorbed by the Z5 zeolite per unit  mass as a function of absolute pressure applied: (a) BK volume, (b) total volume adsorbed up to complete saturation. The pressure values on the x axes of the two plots are different because in (a) the average is evaluated only considering the pressure measurement up to the BK, while in (b) the whole run was considered. The highest pressure point in plot (a) has been excluded due to instabilities in the first phase of the run, which resulted in an unreliable measurement. Both graphs include results obtained using GC PPU and MS columns. The \SI{270}{g} data were linearly fitted using Tukey robust regression~\cite{regression}. The accuracy of the regression model is estimated through the Median Absolute Error (MedAE). Results of two runs performed with \SI{2}{kg} of Z5 zeolite are superimposed for comparison (in (b) the PPU points are below the MS one).}
\label{fig:V_PerMassUnit_Vs_Pressure}
\end{figure}

From the results presented in Fig.~\ref{fig:V_PerMassUnit_Vs_Pressure}, it was observed that increased pressure significantly enhances the methane adsorption capacity of Z5. Measurements recorded at $p > \SI{5}{bar}$ were compromised by pressure leaks during the measurements. These leaks were attributed to the limited sealing capabilities of the O-ring connecting the cartridge to the gas line under higher pressures. Despite these experimental challenges, the positive correlation between pressure and methane adsorption remains evident, suggesting that pressurizing the gas collected from the barn would substantially improve system performance.

\subsection{Evaluation of adsorption using a higher mass}
To minimize the frequency of regeneration cycles upon zeolite saturation, the final prototype will employ cartridges containing a significantly larger mass of zeolite $(2.0\pm \SI{0.2}{kg}),$ thereby increasing the total methane adsorption capacity.

Two runs were taken with these larger cartridges, using an experimental setup identical to the one depicted in the previous tests. 
The volume of methane adsorbed per unit mass of Z5 zeolite was evaluated under these new conditions, with the results presented in Fig.~\ref{fig:V_PerMassUnit_Vs_Pressure}, together with the results obtained with the \SI{270}{g} cartridges as reference.

A comparison between the specific adsorption volumes (methane adsorbed per unit mass) of the \SI{0.27}{kg} cartridges and the \SI{2}{kg} cartridges reveals a significant discrepancy. The data indicate that the specific amount of \ce{CH4} adsorbed for the larger Z5 cartridge, in the two different measurements performed at a pressure around 1-1.2 bar, was 46\% (at p$\sim$\SI{1}{bar}) and 73\% (at p$\sim$\SI{1.2}{bar}) of that in similar conditions (p$\sim$\SI{1.1}{bar}). For the first run at lower pressure, an anomaly during the TSA regeneration process is suspected. 
However, the reduction in specific adsorption is observed also in the second run, which involved  a longer data collection period, indicating that it may be symptomatic of a geometric effect caused by non-uniform  methane flow and adsorption upon entering the larger cartridge, as explained in Ref.~\cite{Tesi_Francesco}. Further data collection is required to substantiate this hypothesis.

\subsection{Results summary}
The tests conducted in the laboratory  highlighted the following aspects:
\begin{enumerate}
    \item It is possible to adsorb \ce{CH4} using commercial zeolites. Zeolites Z5 (\SI{5}{\angstrom}) and Z10 (13X) are the most effective commercially available zeolites for capturing \ce{CH4}.
    \item Both regeneration processes, Thermal Swing Adsorption (TSA) and Vacuum Swing Adsorption (VSA), are effective for regenerating zeolites saturated with \ce{CH4}. VSA is more energy-efficient than TSA, thus zeolite Z5 was ultimately selected as it is regenerated more effectively with this method.
    \item Scalability: the adsorption process is effective both on small volumes (the test cartridges contain approximately \SI{270}{g} of adsorbent material) and on larger volumes, in the order of \SI{2}{kg}.
    \item Pressure effect: it has been demonstrated that the adsorption capacity increases linearly with pressure (tested up to \SI{5}{bar}). Physically, the process is a balance between mechanical capture in the pores of the zeolite and thermal agitation, which promotes the release of molecules. Increasing the pressure shifts this balance in favor of capture.
\end{enumerate}

\section{Design of the \ce{CH4} recuperation system}
In this section, the methane recuperation system to be installed in the barn is presented.

\begin{figure}[h]
    \centering
    \includegraphics[width=\textwidth]{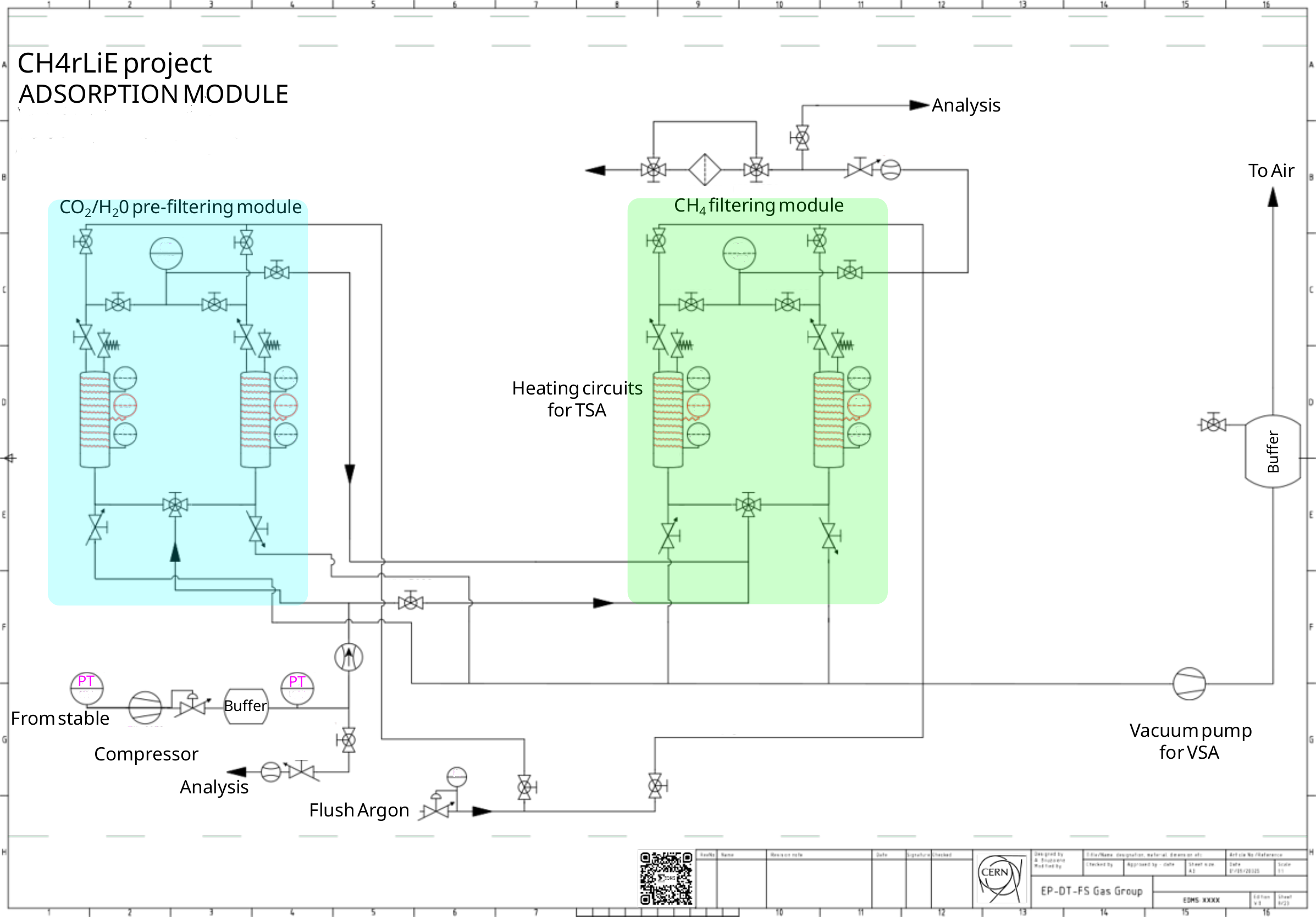}
    \caption{Scheme of recuperation system to be installed in the barn. }
    \label{fig:prototipoFinal}
\end{figure}

The prototype, illustrated in Fig.~\ref{fig:prototipoFinal}, consists of three main components arranged in series: 
\begin{itemize}
    \item a pump to extract air from the barn and increase the gas pressure within the system (maximum gauge pressure $p = \SI{3}{bar}$). This exploits the positive correlation between pressure and the methane adsorption capacity of the materials;
    \item a pair of cartridges in parallel, each with a mass of \SI{2}{kg}, designed to selectively adsorb humidity and  \ce{CO2} from the barn air. These cartridges are filled with a 50:50 mixture of Z3 (for moisture adsorption) and Z4 (for the \ce{CO2}). The selection of these zeolites for the pre-filtering stage has been done based on the results in Ref.~\cite{Tesi_MCA};
    \item in series with the aforementioned components, two Z5 cartridges (with the same mass of \SI{2}{kg} and in the same configuration). These are specifically dedicated to methane adsorption.
\end{itemize}

To monitor the composition of the output gas, an analytical station equipped with \ce{CO2} and \ce{CH4} sensors can be connected to the exhaust line.
The operating logic requires the cartridges to be brought to saturation one at a time. Once both cartridges in one of the filtering stages are saturated, the capture operation is suspended to regenerate each cartridge individually before restarting the cycle.

To regenerate the saturated cartridges on-site, the system is equipped with lines capable of performing both VSA regeneration and the TSA regeneration. The TSA method serves as a secondary option for initial conditioning and in scenarios where VSA proves ineffective.  Eventually, the \ce{CH4} extracted during regeneration can be stored in a buffer for possible later reuse.

\section{Evaluation of the impact of the project}

Assuming a baseline concentration of \ce{CH4} of approximately \SI{100}{ppm} within the livestock facility environment, roughly \SI{100}{kg} of commercial Z5 zeolite adsorbent is required to capture \SI{1}{\liter} of \ce{CH4}.
The current prototype utilizes cartridges containing \SI{2}{kg} of Z5, operating in an alternating sequence. Consequently, at full saturation and an operating pressure of \SI{1}{bar}, the system achieves a maximum capture capacity of \SI{40}{\milli\liter} of \ce{CH4}. Increasing the operating pressure to \SI{3}{bar} yields a proportional increase, adsorbing approximately \SI{120}{\milli\liter} of \ce{CH4}.
Given that a single dairy cow emits approximately \SI{164.9}{\kilo\liter} (or \SI{164.9}{\cubic\meter}) of \ce{CH4} annually, averaging \SI{450}{L} per day at \SI{20}{\celsius}, the present prototype does not yet have a substantial environmental or economic impact. However, it successfully establishes a proof-of-concept for a highly promising technology.
System performance could be significantly enhanced through the following modifications:
\begin{itemize}
    \item Scaling the total mass of the Z5 zeolite adsorbent to \SI{100}{kg}, distributed across a battery of twenty \SI{5}{kg} cartridges.
    \item  Elevating the operating pressure up to \SI{10}{bar}.
\end{itemize}

Assuming the linear correlation between adsorption capacity and gas pressure extends beyond the currently tested experimental range, these modifications would enable the capture of up to \SI{100}{\liter} of \ce{CH4} per operational cycle.
A more substantial optimization could be achieved by replacing the Z5 zeolite with an advanced material possessing a markedly higher \ce{CH4} adsorption capacity, ideally an order of magnitude higher. Such a material substitution would theoretically allow the system to adsorb nearly the entirety of a single animal annual \ce{CH4} emissions.

\section{Conclusions}
The CH4rLiE project has successfully demonstrated the technical feasibility of adapting high-energy physics gas recuperation systems for reducing GHG emissions in the agricultural sector. Laboratory validation of the prototype shows that methane can be effectively captured and recovered even at low concentrations through the integration of specific adsorbent materials. Future work will focus on the field-testing of the full-scale system to evaluate its long-term performance and stability in operational barn environments.

\section*{Funding}
This research was financed by the European Union – Next Generation EU PRIN 2022 PNRR - P2022FTF7L

\section*{Acknowledgements}
The authors would like to sincerely thank the electronics service and the mechanical workshop of the Physics Department of the University of Pavia and of the INFN Sezione di Pavia. Special thanks are also due to the CERN EP-DT Gas Group for providing the facilities and hosting the experimental activities described in this work.

The authors also acknowledge the support of the LIFE CLINMED-FARM project (Towards a Mediterranean Climate Neutral Farm Model, LIFE20 CCM/ES/001751).

\section*{Data availability}
The data that support the findings of this study are available from the corresponding author upon reasonable request.

\section*{Conflicts of Interest} The authors declare that they have no known competing financial interests or personal relationships that could have appeared to influence the work reported in this paper.

\bibliographystyle{elsarticle-num} 
\bibliography{biblio}
\end{document}